\documentclass[a4paper,11pt]{article}
\pdfoutput=1
\usepackage{jcappub}
\usepackage{graphicx,epstopdf,mathrsfs,amssymb,verbatim,color,pifont,longtable,tabu,booktabs}
\usepackage{multirow,amsmath,subfig,slashed,float,sidecap,mathtools,diagbox,array}
\usepackage[dvipsnames]{xcolor}
\usepackage[normalem]{ulem}
\usepackage[parfill]{parskip}
\usepackage{orcidlink}
\newcolumntype{P}[1]{>{\centering\arraybackslash}p{#1}}

\numberwithin{equation}{section}

\title{Dark matter cooling during early matter-domination boosts sub-earth halos}

\author[a]{Avik Banerjee\orcidlink{0000-0003-2475-8978},}
\author[b]{Debtosh Chowdhury\orcidlink{0000-0002-4302-7356},}
\author[b]{Arpan Hait\orcidlink{0000-0001-7540-0111},}
\author[b]{\\ Md\,Sariful Islam\orcidlink{0009-0001-7174-7369
}}

\affiliation[a]{Department of Theoretical Physics, Tata Institute of Fundamental Research, Homi Bhabha Road, Mumbai 400005, India}
\affiliation[b]{Department of Physics, Indian Institute of Technology Kanpur, Kanpur, 208016, India}

\emailAdd{avik.banerjee\_205@tifr.res.in}
\emailAdd{debtoshc@iitk.ac.in}
\emailAdd{arpan20@iitk.ac.in}
\emailAdd{sariful21@iitk.ac.in}

\abstract{The existence of an early matter-dominated epoch prior to the Big Bang Nucleosynthesis (BBN) may lead to a scenario where the thermal dark matter cools faster than plasma before the radiation-dominated era begins. In the radiation-dominated epoch, dark matter free-streams after it decouples both chemically and kinetically from the plasma. In the presence of an early matter-dominated era, chemical decoupling of the dark matter may succeed by a partial kinetic decoupling before reheating ends, depending upon the contributions of different partial wave amplitudes in the elastic scattering rate of the dark matter. We show that the $s$-wave scattering is sufficient to partially decouple the dark matter from the plasma, if the entropy injection during the reheating era depends on the bath temperature, while $p$-wave scattering leads to full decoupling in such cosmological backdrop. The decoupling of dark matter before the end of reheating causes an additional amount of cooling, reducing its free-streaming horizon compared to the usual radiation-dominated cosmology. The enhanced matter perturbations for scales entering the horizon prior to the end of reheating, combined with the reduced free-steaming horizon, increase the number density of sub-earth mass halos. The resulting boost in the dark matter annihilation signatures could offer an intriguing probe to differentiate pre-BBN non-standard cosmological epochs. We show that the free-streaming horizon of the dark matter requires to be smaller than a cut-off to ensure a boost in the sub-earth halo populations. As case studies, we present two examples: one for a scalar dark matter with $s$-wave elastic scattering and the other one featuring a fermionic dark matter with $p$-wave elastic scattering. We identify regions of parameter space in both models where the dark matter kinetically decouples during reheating, amplifying small-scale structure formation.}

\begin{document}
\maketitle
\flushbottom


\section{Introduction}
\label{sec:intro}

A non-luminous form of matter, dubbed as dark matter (DM)~\cite{Zwicky:1933gu}, is believed to constitute almost 85\% of the total matter content and 28$\%$ of the total energy budget of the universe~\cite{Planck:2018vyg}. Several astrophysical~\cite{Zwicky:1933gu, Zwicky:1937zza, Rubin:1970zza, Clowe:2006eq, Massey:2010hh, Hu:2001bc} and cosmological~\cite{Planck:2018vyg} observations substantiate the claim for the existence of DM, leveraging its gravitational interaction. However, little is understood regarding other characteristics of the DM, such as its mass, spin, and potential interactions with visible matter, if any. To exemplify the uncertainties surrounding the mass of DM, it spans from fuzzy DM with a mass approximately $\sim10^{-22}$ eV~\cite{Hu:2000ke, Hui:2016ltb, Hlozek:2014lca, Khmelnitsky:2013lxt, Chowdhury:2023xvy} to the most massive primordial black holes, which are around $\sim 10^{4}$ times the solar mass~\cite{Garcia-Bellido:2017fdg, AlvesBatista:2021eeu}.

Testability of weakly interacting massive particles (WIMPs) through a variety of direct and indirect detection experiments establish them as one of the most promising candidates for dark matter~\cite{Steigman:1984ac, Jungman:1995df, Bertone:2004pz, Steigman:2012nb, Arcadi:2017kky}, in the mass range between a few GeV to TeV. In the WIMP scenario, the DM remains in equilibrium with the thermal plasma of visible matter in the early universe. As the universe expands, after some time the interaction rate ($\Gamma$) of DM with particles in the thermal bath becomes ineffective (i.e., $\Gamma < H$, where $H$ is the Hubble expansion rate) to keep it in thermal equilibrium with the plasma. As a result, the DM thermally decouples from the standard model plasma. The thermal decoupling results in a freeze-out of the DM, and it sets the relic density of DM, which at the present day has a value of $\Omega_{\rm DM}^{0}h^{2}\simeq 0.12$, as measured by the Planck collaboration~\cite{Planck:2018vyg}.

In the WIMP scenario, it is important to note that the thermal decoupling of DM implies both chemical and kinetic decoupling from the plasma. Chemical decoupling of DM sets the number density of relic DM in the universe, however, full thermal decoupling of DM is crucial for forming structures. While chemical decoupling is primarily governed by the annihilation of the DM ($\chi$) into the bath particles ($B$), represented as $\chi\chi \to BB$, the kinetic decoupling depends on the elastic scattering processes, i.e. $\chi B\to \chi B$. In general, chemical and kinetic decoupling may not take place at the same time, since the annihilation rate of the DM $(\Gamma_{\rm ann}= n^{\rm eq}_\chi \langle\sigma v\rangle_{\rm ann})$ and the elastic scattering rate $(\Gamma_{\rm el}= n^{\rm eq}_B \langle\sigma v\rangle_{\rm el})$ vary with the bath temperature in a qualitatively different manner. For a cold dark matter, i.e., if the freeze-out is non-relativistic, the equilibrium number density of the DM after the freeze-out falls as $n^{\rm eq}_\chi\sim T^{3/2}e^{-m_\chi/T}$, where $m_\chi$ is the mass of the DM and $T$ represents the temperature of the bath. In contrast, the relativistic bath particles' number density varies as $n^{\rm eq}_B\sim T^3$. Thus, in a radiation-dominated universe ($H\sim T^2/M_{\rm pl}$), chemical decoupling precedes the kinetic decoupling of cold dark matter, assuming that the thermal averaged cross-section mildly varies with temperature. 

In the standard lore, the DM is presumed to undergo decoupling in a universe dominated by radiation prior to the commencement of Big Bang Nucleosynthesis (BBN). Nevertheless, hardly any evidence exists to confirm that the universe was only radiation-dominated prior to the BBN, since the end of inflation. This ambiguity prompts various conjectures regarding the alternative thermal evolution of the cosmos, a pre-BBN early matter-dominated epoch (EMDE) emerging as a prominent hypothesis.  Several motivated scenarios beyond the Standard Model (BSM) propose the existence of meta-stable oscillating scalar fields, such as string theory inspired moduli fields~\cite{Moroi:1999zb, Starobinsky:1994bd, Dine:1995uk}, supersymmetric condensates and gravitino~\cite{Thomas:1995ze, Moroi:1994rs}, scenarios with primordial black hole formation~\cite{Khlopov:1980mg, Khlopov:2008qy, Ketov:2019mfc}, inflaton fields~\cite{Allahverdi:2002nb, Allahverdi:2010xz}, curvaton~\cite{Moroi:2002rd}, dilaton~\cite{Lahanas:2011tk}, and Q-balls~\cite{Fujii:2002kr}, among others. These long-lived species could instigate an early matter-dominated epoch after inflation, followed by reheating due to the late decay of these oscillating matter fields into radiation~\cite{Albrecht:1982mp, Turner:1983he, Kofman:1994rk, Shtanov:1994ce, Kofman:1997yn, Bassett:2005xm}. The presence of such an EMDE strongly impacts both the thermal and non-thermal DM production~\cite{Chung:1998rq, Giudice:2000ex, Allahverdi:2002pu, Mukaida:2012bz, Harigaya:2014waa, Choi:2015yma, Mambrini:2016dca, Garcia:2017tuj, Hamdan:2017psw, Drees:2017iod, Maity:2018dgy, Bernal:2018qlk, Maity:2018exj, Bhattacharyya:2018evo, Garcia:2018wtq, Drees:2018dsj, Chowdhury:2018tzw, Maldonado:2019qmp, Harigaya:2019tzu, Banerjee:2019asa, Arias:2020qty, Garcia:2020eof, Haque:2020zco, Bhatia:2020itt, Ballesteros:2020adh, Garcia:2020wiy, Mambrini:2021zpp, Drees:2021lbm, Garcia:2021gsy, Haque:2021mab, Clery:2021bwz, Banerjee:2022fiw, Bernal:2022wck, Haque:2022kez, Chowdhury:2023jft, Becker:2023tvd, Silva-Malpartida:2023yks, Bernal:2023ura, Barman:2024mqo, Bernal:2024yhu} and their decoupling~\cite{Gelmini:2008sh, Visinelli:2015eka, Waldstein:2016blt}. 
 
The kinetic decoupling of DM during the early matter-domination (EMD) is determined by how the elastic scattering cross-section of the DM and the Hubble expansion rate of the universe varies with the plasma temperature during reheating.  
Reheating initiates once the injection of entropy from the meta-stable scalar field ($\phi$), which dominates the energy density of the universe during the EMDE, into the plasma becomes prominent, i.e., when the dissipation rate of the long-lived $\phi$ surpasses the Hubble expansion ($\Gamma_\phi > H$) rate. The expansion rate of the universe during EMDE is contingent upon the dissipation rate of the scalar field. For instance, if $\Gamma_\phi$ is constant, entropy injection leads to a bath of thermalized particles, where the bath temperature scales with scale factor as $T \propto a^{-3/8}$ and the Hubble expansion rate scales as $H \propto T^{4}$~\cite{Giudice:2000ex}. 

In such an EMDE scenario, the $s$-wave elastic scattering of non-relativistic DM ($\langle\sigma v\rangle_{\rm el}\sim {\rm const.}$) is insufficient to kinetically decouple the DM from the plasma before the onset of radiation-dominated epoch, as the momentum exchange rate in elastic scattering ($\gamma_{\rm el}=\Gamma_{\rm el}T/m_\chi$) and the Hubble expansion both scale as $T^4$. Thus, the scales entering the horizon prior to the end of reheating can not lead to the formation of protohalos~\cite{Barenboim:2013gya, Barenboim:2021swl, Delos:2021rqs}. However, $p$-wave elastic scattering ($\langle\sigma v\rangle_{\rm el}\sim T^2$) yielding $\gamma_{\rm el}\sim T^6$ leads to a partial kinetic decoupling, as shown in~\cite{Waldstein:2016blt}. Consequently, the DM begins to cool at a faster rate than the plasma during the EMDE. Eventually, after reheating, DM fully decouples from the plasma. Due to this partial decoupling, the additional cooling of DM during EMDE reduces the free-streaming horizon compared to the standard scenario of DM decoupling in the radiation-dominated epoch. Thus, the free-streaming of DM in this type of non-standard scenario does not suppress small-scale structures (sub-earth halos) formed by the scales entering the horizon before reheating. On the contrary, such small scales are additionally endowed with enhanced matter perturbations in the presence of an EMDE causing the formation of enhanced structures compared to the standard scenario \cite{Erickcek:2011us, Fan:2014zua, Das:2021wad}.

In this paper, we consider an alternate proposition where the entropy injection during the EMDE depends on the temperature of the bath (i.e., $\Gamma_\phi\sim T$). This situation arises when the back-reaction of the thermal decay products of the oscillating scalar field $\phi$ leads to a dominant temperature-dependent contribution to the dissipation rate $\Gamma_\phi$~\cite{Mukaida:2012qn}. Compared to the $\Gamma_\phi\sim {\rm const.}$ scenario, temperature and the Hubble expansion rate scales differently, namely $T\propto a^{-1/2}$ and $H\propto T^3$, during the EMDE with $\Gamma_\phi\sim T$~\cite{Banerjee:2022fiw}. We will demonstrate that in such a scenario, $s$-wave elastic scattering of DM with the bath particles is enough to partially decouple the DM before the end of reheating. In this alternate thermal history, $p$-wave elastic scattering of DM leads to a full decoupling of the DM during the EMDE. Consequently, the extra cooling of the DM from its thermal decoupling till the reheating, and the enhanced matter perturbations during EMDE boost the formation of structures at sub-earth scales. We further compute the enhanced density perturbations of the DM during such EMDE and the resulting boost in the number density of sub-earth halos produced during this epoch.  

As a proof of principle, we consider two simple models of dark matter, one with a scalar DM and the other with a fermionic DM, where the DM elastically scatters off radiation through $s$-wave and $p$-wave partial wave amplitude channels, respectively. For both cases, we demonstrate that the parameter space that satisfies the observed relic density can also lead to kinetic decoupling of the DM during the EMDE yielding a reduced free-streaming horizon of the DM compared to purely radiation-dominated cosmology. Additionally, we demonstrate that, despite kinetic decoupling during the EMDE, the enhancement in small-scale structure formation occurs only within a specific region of parameter space where the free-streaming horizon falls below a certain cut-off value.

The formation of microhalos may lead to an enhanced annihilation rate of DM, providing an intriguing probe for pre-BBN thermal history. The detection of gamma rays from DM annihilation at the galactic center by the indirect detection experiments~\cite{Gondolo:1999ef, Hooper:2010mq, IceCube:2015rnn}, such as the Fermi-LAT~\cite{Fermi-LAT:2015att, MAGIC:2016xys, Fermi-LAT:2017opo} place strong constraints on the DM annihilation rate. Ref.~\cite{Choi:2017ncz} further places a constraint on the reheating temperature from the WIMP dark matter annihilation using the Fermi-LAT data, assuming a fixed mass for the microhalos. In the presence of an EMDE, two compensatory effects determine the overall annihilation rate of DM: (i) a reduced annihilation cross-section of thermal DM to match the correct relic density due to entropy dilution, and (ii) a boost in the number density of microhalos resulting from enhanced matter perturbations and additional cooling of partially decoupled DM before reheating. While we show the enhancement of microhalo population due to the presence of an EMDE with temperature-dependent entropy injection, a full dedicated study to compute the boost factor for DM annihilation rate signatures is deferred to a future work~\cite{Banerjee:202xabc}. Besides annihilation signatures of DM~\cite{Bringmann:2011ut, Erickcek:2015bda, StenDelos:2019xdk, Blanco:2019eij, Ganjoo:2024hpn}, sub-halo collisions with old neutron stars, sub-halo dark matter scatterings with cosmic rays~\cite{Bramante:2021dyx}, pulsar timing arrays~\cite{Siegel:2007fz, Baghram:2011is, Choi:2017ncz, Dror:2019twh}, microlensing~\cite{Blinov:2021axd, Bechtol:2022koa}, and gravitational waves produced in EMDE~\cite{Ballesteros:2022hjk, Barman:2022qgt, Barman:2023rpg, Barman:2023ktz, Bernal:2023wus, Xu:2024fjl, Soman:2024zor} could offer alternative methods to investigate non-standard cosmological epochs, due to an increased sub-earth halo population, as described in this paper.

The paper is organized as follows. In Sec.~\ref{partial_decoup} we discuss the mechanism of kinetic decoupling in different cosmological backdrops and compute the free-streaming horizon of the DM. In Sec.~\ref{perturbations}, we first describe the thermal history of an EMDE with temperature-dependent entropy injection. The rest of the section is dedicated to a detailed study of the cosmological perturbations during the EMDE and the formation of sub-earth mass halos. Two simple models featuring a scalar and a fermionic dark matter which exhibit $s$-wave and $p$-wave elastic scatterings, respectively, and decoupling during the EMDE are explored as case studies in Sec.~\ref{DM_EMDE}. We conclude in Sec.~\ref{conc}.

\section{Partial kinetic decoupling of dark matter}
\label{partial_decoup}


We briefly review the general formalism to study the decoupling of DM in different thermal histories of the universe. We assume that the DM chemically decouples from the thermal bath at the freeze-out temperature ($T_{\text{fo}} \sim m_{\chi}/20$), while still remaining kinetically coupled to the plasma. The Boltzmann equation governing the evolution of the phase space distribution of the DM ($f_\chi({\mathbf p},t)$) is given by~\cite{Bernstein:1988bw, Kolb:1990vq}
\begin{align}
	E \left(\partial_{t} - H {\mathbf p}\cdot {\mathbf \nabla_{{\mathbf p}}} \right) f_{\chi}({\mathbf p}, t) = \mathcal{C}_{\rm ann}[f_{\chi}] + \mathcal{C}_{\rm el}[f_{\chi}]\,,
 \label{BEQ:Phase_DM}
\end{align}
where $\mathcal{C}_{\rm ann}[f_{\chi}]$ and $\mathcal{C}_{\rm el}[f_{\chi}]$ denote the collision terms corresponding to the annihilation ($\chi\chi\to BB$) and the elastic scatterings ($\chi B\to \chi B$) of the DM with the bath particles ($B$), respectively. The number density ($n_\chi$) and the temperature ($T_\chi$) of the DM is defined by taking the moments of the phase space distribution as~\cite{Bringmann:2006mu, Visinelli:2015eka, Bringmann:2016ilk}
\begin{eqnarray}\label{EQ:DM_Temp}
	n_{\chi}(t) \equiv g_{\chi} \int \frac{d^3 p}{(2\pi)^3} f_{\chi}({\mathbf p},t)\,, \quad {\rm and} \quad T_{\chi}(t) \equiv \frac{g_{\chi}}{3 n_{\chi}} \int \frac{d^3 p}{(2\pi)^3} \frac{p^{2}}{E} f_{\chi}({\mathbf p},t)\,,
\end{eqnarray}
where $p\equiv |{\mathbf p}|$ and $g_\chi$ denotes the number of internal degrees of freedom of $\chi$. The evolution of $n_\chi(t)$ is obtained by integrating both sides of Eq.\,\eqref{BEQ:Phase_DM} as
\begin{align}
     \frac{dn_{\chi}}{dt} + 3 H n_\chi = g_\chi \int \frac{d^3 p}{(2\pi)^3} \frac{\mathcal{C}_{\rm ann}[f_{\chi}]}{E} = - \langle \sigma v \rangle_{\rm ann} (n^2_\chi - n_{\chi,{\rm eq}}^2)\,,
\end{align}
where $n_{\chi,{\rm eq}}$ denotes the equilibrium number density of the DM. The integration over $C_{\rm el}[f_\chi]$ vanishes since the elastic scattering does not change the number of DM. The velocity averaged annihilation cross-section of the DM into bath particles is given by~\cite{Kolb:1990vq, Gondolo:1990dk} 
\begin{eqnarray}\label{Eq:th_avg_crss}
	\langle \sigma v \rangle_{\rm ann}  = \frac{1}{8 m_{\chi}^{4}T K_{2}\left(m_{\chi}/T\right)^{2}} \int_{4m_{\chi}^{2}}^{\infty} ds\, \sqrt{s} \left(s-4m_{\chi}^{2}\right)\sigma_{\rm ann}(s)K_{1}\left(\sqrt{s}/T\right)\,,
 \label{eq:sigma_ann}
\end{eqnarray}
where $K_{1,2}$ are the modified Bessel function of the second kind. Once the universe cools down to $T_{\rm fo}$,  DM leaves the chemical equilibrium, and the co-moving number density of the DM freezes out.

At a temperature $T$ after chemical decoupling of the DM, the annihilation rate becomes negligible compared to the Hubble expansion rate. So, one can drop $\mathcal{C}_{\rm ann}[f_{\chi}]$ from Eq.~\eqref{BEQ:Phase_DM}. The collision term corresponding to the elastic scattering of the non-relativistic DM  has the form~\cite{Binder:2017rgn}  
\begin{eqnarray}\label{EQ_Elastic_Scattering_rate}
\mathcal{C}_{\rm el}[f_{\chi}] = \frac{E}{2}\gamma_{\rm el}(T)\left[T \frac{\partial}{\partial p}\left(E \frac{\partial f_{\chi}}{\partial p}\right) + \frac{\partial}{\partial p}\left(p f_{\chi}\right) \right].
\end{eqnarray}
The momentum transfer rate in the elastic scattering $\gamma_{\rm el}(T)$ is given by~\cite{Bringmann:2006mu, Bringmann:2016ilk}
\begin{align}
	\gamma_{\rm el}(T) 
	= \frac{1}{48\pi^3 T m_{\chi}^{3}}\int d\omega\, k^{4}g^{\pm}(\omega)[1 \mp g^{\pm}(\omega)] \left\langle \mathcal{M}^{2} \right\rangle_{t}\,,
\qquad \left \langle \mathcal{M}^{2}\right\rangle_{t} \equiv \frac{1}{8k^{4}} \int_{-4k^{2}}^{0}dt (-t)\mathcal{M}^{2}\,.
\end{align}
Here $(\omega, {\mathbf k})$ represents the momentum of the incoming bath particle, $k\equiv |{\mathbf k}|$ and $\mathcal{M}^2$ is the squared amplitude of the elastic scattering process, while $g^{\pm}(\omega)=1/[\exp(\omega/T)\pm 1]$ denote the distribution functions of the relativistic bath particles. For $s$-wave elastic scatterings $\langle \mathcal{M}^{2}\rangle_{t} = {\rm const.}$, while for $p$-wave $\langle \mathcal{M}^{2}\rangle_{t} \propto \omega^2$. After chemical decoupling, the evolution of the temperature of the DM, i.e., $T_\chi$ with respect to the scale factor is obtained from Eq.~\eqref{BEQ:Phase_DM} as~\cite{Bringmann:2006mu, Waldstein:2016blt}
\begin{eqnarray}\label{eq:evl_DM_Temp}
	\frac{dT_{\chi}}{d\ln{a}} + 2 T_{\chi} (a) \left[ 1 + \frac{\gamma_{\rm el}(a)}{H(a)}\right] = 2 \frac{\gamma_{\rm el}(a)}{H(a)} T(a).
\end{eqnarray} 
To arrive at this equation, we specifically assume that the DM is non-relativistic below the freeze-out temperature ($m_{\chi} > T_{\rm fo}$). As long as the DM is thermally coupled to the plasma, $T_\chi$ is equal to the temperature of the plasma ($T$). When the elastic scattering rate becomes smaller than the Hubble expansion rate i.e., $\gamma_{\rm el}(T) \ll H(T)$, the solution of Eq.~\eqref{eq:evl_DM_Temp} shows that $T_{\chi}$ redshifts as $T_{\chi}\sim a^{-2}$. This solution additionally necessitates another condition, i.e., $\gamma_{\rm el}(T) T\ll H T_\chi$. We will explore the implications of relaxing this condition in the subsequent sections. In a radiation-dominated universe, the bath temperature evolves as $T\sim a^{-1}$ while the DM temperature cools as $T_{\chi}\sim a^{-2}$ after kinetic decoupling. However, the situation becomes more involved if the thermal history of the universe during the decoupling is different from the usual radiation domination~\cite{Gelmini:2008sh, Visinelli:2015eka, Waldstein:2016blt}. In the following, we investigate the kinetic decoupling of the DM for non-standard cosmological backdrops.

\subsection{Kinetic decoupling in non-standard cosmological scenario}
\label{sec:kin_decoup}

We consider the scenarios where the temperature of the thermal bath evolves with the scale factor as $T \propto a^{-\alpha}$ with parameter $\alpha\geq 0$, and the Hubble expansion rate varies with temperature as $H\propto T^\beta$. The momentum transfer rate in the elastic scattering depends on the temperature of the bath as~\cite{Visinelli:2015eka, Waldstein:2016blt}
\begin{eqnarray}
	\gamma_{\rm el}(T) \propto T^{(4+n)}\,,
\end{eqnarray}
where $n=0, 2, ...$ for $s$-wave, $p$-wave, and so on. The general solution to the Eq.~\eqref{eq:evl_DM_Temp} has the form~\cite{Visinelli:2015eka}\footnote{$\Gamma(q, s)$ is the upper incomplete gamma function: $\Gamma(q, s)$ = $\int_{s}^{\infty} dt\, t^{q-1}e^{-t}$.}
\begin{eqnarray}\label{eq:Temp_DM_full}
  T_{\chi}(a) = T \left(\frac{s}{s_{\rm dec}} \right)^{\lambda}e^{s-s_{\rm dec}} + T\,s^{\lambda}\,e^{s}\left[\Gamma(1-\lambda, s) - \Gamma(1-\lambda, s_{\rm dec})\right],
\end{eqnarray}
\begin{align}
    {\rm with} \quad \lambda = \frac{(2-\alpha)}{\alpha(4+n-\beta)}\,, \quad {\rm and} \quad s = \frac{2}{\alpha(4+n-\beta)} \frac{\gamma_{\rm el}(T)}{H(T)}\left(\frac{T}{T_{\rm dec}}\right)^{(4+n-\beta)}\,.  
\end{align}

We define $T_{\rm dec}$ ($a_{\rm dec}$) as the temperature (scale factor) when $\gamma_{\rm el}(T_{\rm dec}) = H(T_{\rm dec})$.
As indicated earlier, the condition for kinetic decoupling, $\gamma_{\rm el}(T)\ll H$, does not necessarily imply that the right-hand side of Eq.~\eqref{eq:evl_DM_Temp} is also negligible compared to $T_\chi$. In fact, the approximate solution of Eq.~\eqref{eq:evl_DM_Temp} in the limit $\gamma_{\rm el}(T)\ll H$, but $\gamma_{\rm el}(T) T \nless H T_\chi$ is given by
\begin{eqnarray}\label{eq:Temp_DM}
	T_{\chi}(a) \simeq \frac{T_{\rm dec}}{2 - \alpha(5+n-\beta)}\left[  2\left(\frac{a}{a_{\rm dec}} \right)^{-\alpha(5+n-\beta)} -\alpha(5+n-\beta) \left(\frac{a}{a_{\rm dec}} \right)^{-2} \right].
\end{eqnarray}
Clearly, the first term in the Eq.~\eqref{eq:Temp_DM} dominates at $a \gg a_{\rm dec}$,  iff $\alpha\left(5+n-\beta\right)< 2$. The necessary requirement to satisfy the first decoupling condition, i.e.,  $\gamma_{\rm el}(T)\ll H$ is $n >n_{\rm dec} = \left(\beta - 4\right)$. When both conditions $n > n_{\rm dec}$ and $\alpha(5+n-\beta)< 2$ are met simultaneously, dark matter cools faster than the plasma, albeit not as rapidly as it would if $T_\chi\sim a^{-2}$. This stage of cooling leads to a partial kinetic decoupling of the DM from the plasma~\cite{Waldstein:2016blt}. We summarize the different possibilities below
$$
\centering
\begin{tabular}{rll}
$n \leq n_{\rm dec}$: & & no kinetic decoupling, \\
$n_{\rm dec} < n < n_{\rm partial}$: & & partial kinetic decoupling, \\
$n >n_{\rm dec}$ and $n \geq n_{\rm partial}$: & & full kinetic decoupling,
\end{tabular}
$$
where $n_{\rm partial}\equiv (2/\alpha)+\beta-5$. Hence, the kinetic decoupling of dark matter significantly depends on the specific dark matter model via the nature of elastic scattering, as well as on the underlying cosmological model through the thermal evolution of the plasma.

\subsubsection*{Partial decoupling during entropy injection}

Now we focus on a scenario where a meta-stable scalar field $\phi$ with an equation of state $\omega_\phi$ dominates the energy density of the universe prior to the onset of BBN. The field $\phi$ coherently oscillates around the minima of its potential and dissipates its energy via decaying into radiation. We assume that the decay products of $\phi$ instantaneously thermalize\footnote{See~\cite{Harigaya:2014waa, Mukaida:2015ria, Garcia:2018wtq, Harigaya:2019tzu, Mukaida:2022bbo, Chowdhury:2023jft, Mukaida:2024jiz} for the effects of non-instantaneous thermalization on dark matter production during reheating.} to produce a thermal bath. In general, the dissipation rate of $\phi$ has non-trivial dependence on the scale factor and the temperature of the plasma as~\cite{Mukaida:2012qn, Co:2020xaf,  Banerjee:2022fiw}
\begin{eqnarray}
	\Gamma_{\phi} \propto a^{k} T^{m}.
\end{eqnarray} 
The Boltzmann equations governing the evolution of the energy density of $\phi$ ($\rho_\phi$) and the radiation ($\rho_\gamma$) are given by
\begin{align}
	\frac{d\rho_{\phi}}{dt} + 3(1 + \omega_\phi)H\rho_{\phi} = - (1 + \omega_\phi)\Gamma_{\phi}\rho_{\phi}\,, \quad
	\frac{d\rho_{\gamma}}{dt} + 4H\rho_{\gamma} =  (1 + \omega_\phi)\Gamma_{\phi}\rho_{\phi}\,.
\end{align}
The approximate solution of these equations yields
\begin{align}
    H \propto a^{-3(1 + \omega_\phi)/2}\,, \quad {\rm and} \quad T \propto a^{\frac{2k -3(1+\omega_\phi)}{2(4-m)}}\,.
\end{align}
This implies, in our notation, $\alpha = (3(1+\omega_\phi)-2k)/2(4-m)$ and $\beta = 3(1 + \omega_\phi)(4-m)/\left[3(1+\omega_\phi) - 2k\right]$.
Thus, the conditions on the elastic scattering channel $n$ for the decoupling of DM are
\begin{eqnarray}
	n_{\rm dec} = \frac{3m(1+\omega_\phi)-8k}{2k-3(1+\omega_\phi)}\,, \quad {\rm and} \quad n_{\rm partial} = \frac{(7+3\omega_\phi)(1+m)-10(2+k)}{2k-3(1+\omega_\phi)}\,.
\end{eqnarray}
\begin{table}[h!]
	\begin{center}
		\begin{tabular}{ccccccccccccc}
            \toprule
		\multirow{1}{2cm}{\centering $\phi$ domination} & \multirow{2}{*}{\centering $k$} & \multirow{2}{*}{\centering $m$} & \multirow{2}{*}{\centering $\alpha$} & \multicolumn{4}{c}{Conditions for kinetic decoupling} \\
		\cmidrule{5-8}
            & & & & $n_{\rm dec}$ & $n_{\rm partial}$ & $s$-wave & $p$-wave \\
		\midrule
		\multirow{2}{2cm}{\centering $\omega_\phi=0$ (Matter)}	& 0 & 0 & 3/8 & 0 & 13/3 & -- & partial \\
			 & 0 & 1 & 1/2 & -1 & 2 & partial & full  \\ 
            \midrule 
		\multirow{3}{2cm}{\centering $\omega_\phi=1/3$ (Radiation)} & -1 & 0 & 3/4 & -4/3 & 1/3 & partial & full \\
			 & 1 & 0 & 1/4 & 4 & 11 & -- & --  \\ 
			 & 1 & 2 & 1/2 & 0 & 3 & -- & partial \\ 
            \midrule 
		\multirow{3}{2cm}{\centering $\omega_\phi=1$ (Kination)} & 0 & 0 & 3/4 & 0 & 5/3 & -- & full \\
			 & 0 & 1 & 1 & -1 & 0 & full & full  \\
            & 1 & 1 & 2/3 & 1/2 & 5/2 & -- & partial  \\
            \bottomrule
		\end{tabular}
		\caption{\sf\it Conditions for kinetic decoupling of the DM are shown for different non-standard cosmological scenarios with entropy injection. For the purpose of illustration, some benchmark values of $k$ and $m$ are chosen for three different values of $\omega_\phi=0,1/3,1$, covering various possibilities of kinetic decoupling. In the last two columns, we specifically consider the cases where  $s$-wave ($n=0$) and $p$-wave ($n=2$) scatterings are dominant.}
		\label{table:QD_poss}
	\end{center}
\end{table}

In Table~\ref{table:QD_poss}, we illustrate the nature of kinetic decoupling for some example scenarios with different equation of states and time dependence of dissipation rates of $\phi$. As highlighted in the Introduction, for an EMDE ($\omega_\phi=0$) with a constant dissipation rate ($\Gamma_\phi={\rm const.}$), $s$-wave ($n=0$) elastic scatterings of DM does not kinetically decouple the DM from the plasma, while $p$-wave ($n=2$) scatterings can only partially decouple the DM. On the other hand, if the dissipation rate (aka entropy injection) is proportional to the temperature ($\Gamma_\phi\propto T$), $s$-wave scatterings are enough to partially decouple the DM during the EMDE while $p$-wave scatterings can fully decouple the DM from the thermal plasma.

\subsection{Free-streaming of partially decoupled dark matter}
\label{freestream}

Once chemically and kinetically decoupled from the thermal bath, the dark matter starts to free-stream with a velocity $v_\chi(a)\propto \sqrt{T_\chi(a)}$. The free-streaming horizon $\lambda_{\rm fsh}$ of the DM sets the scale below which the structures will be washed out due to the large DM velocity. The scaling of $T_\chi(a)$ with the scale factor($a$) plays a crucial role in determining $\lambda_{\rm fsh}$. We compute $\lambda_{\rm fsh}$ in two cases, namely, in the presence of an EMDE, and in a purely radiation-dominated (RD) universe using the general solution Eq.~\eqref{eq:Temp_DM_full} for $T_\chi(a)$ and the following expressions
\begin{align}\label{Eq:eq_fsh}
	\lambda^{\rm EMD}_{\rm fsh} & = \int_{t_{\rm dec}}^{t_{0}}dt\, \frac{v_{\chi}(t)}{a(t)} = \sqrt{\frac{3}{m_{\chi}}} \left[ \int_{a_{\rm dec}}^{a_{\rm RH}} + \int_{a_{\rm RH}}^{a_{\rm eq}} + \int_{a_{\rm eq}}^{a_{0}} \right] da \frac{\sqrt{T_{\chi}(a)}}{a^{2}H(a)}\,, && \begin{cases}
	    T\sim a^{-\alpha},\\ H\sim T^\beta \nonumber
	\end{cases}\\ \nonumber\\
 \lambda^{\rm RD}_{\rm fsh} & = \int_{t_{\rm kds}}^{t_{0}}dt\, \frac{v_{\chi}(t)}{a(t)} = \sqrt{\frac{3}{m_{\chi}}} \left[ \int_{a_{\rm kds}}^{a_{\rm eq}} + \int_{a_{\rm eq}}^{a_{0}} \right] da\, \frac{\sqrt{T_{\chi}(a)}}{a^{2}H(a)}\,, && \begin{cases}
	    T\sim a^{-1},\\ H\sim T^2\,.
	\end{cases}
\end{align}

\begin{figure}[t!]
\centering
{\includegraphics[width=\textwidth]{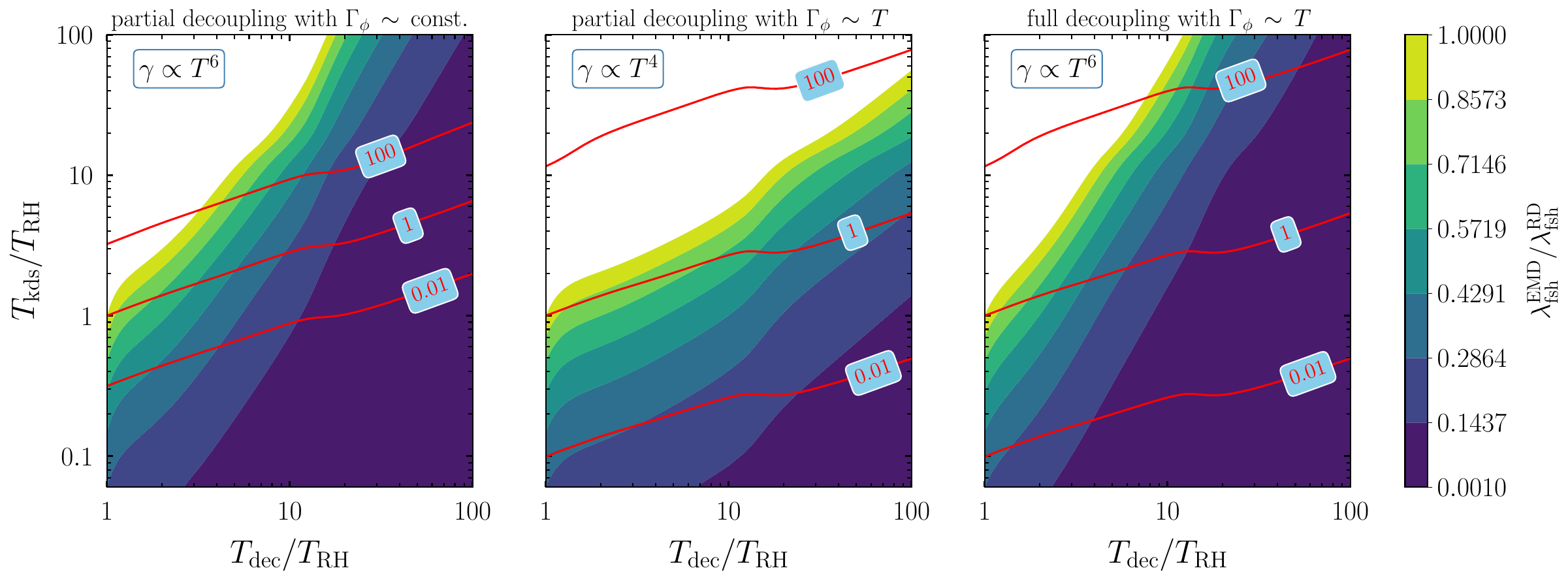}}
\caption{\sf\it The ratios of dark matter free-streaming horizon in the presence of an EMDE and in standard radiation dominated epoch are shown in $T_{\rm kds}-T_{\rm dec}$ plane in the units of $T_{\rm RH}$. We fix the value of $T_{\rm RH}$ at 10 MeV for illustration. The red lines represent the isocontours for the ratio $r$. For $r<1$, the free-streaming horizon in an EMDE is always smaller than its counterpart in RD era. The change in the slope of the isocontours of $r$ around $T_{\rm dec}/T_{\rm RH} \sim 20$ is due to the jump in $g_{*}(T)$ around the QCD phase transition.}
 \label{fig:free_stream}
\end{figure} 

We define $T_{\rm kds}$ as the temperature at kinetic decoupling in the radiation-dominated epoch, such that $\gamma_{\rm el}(T_{\rm kds}) = H_{\rm RD}(T_{\rm kds})$, while $T_{\rm dec}$ corresponds to the decoupling temperature in the EMDE when $\gamma_{\rm el}(T_{\rm dec}) = H_{\rm EMD}(T_{\rm dec})$.

In Fig.~\ref{fig:free_stream}, we present the region in the $T_{\rm kds}-T_{\rm dec}$ plane (normalized to $T_{\rm RH}$), where the ratio $\lambda^{\rm EMD}_{\rm fsh}/\lambda^{\rm RD}_{\rm fsh} < 1$. The solid red lines denote the isocontours of the ratio $r$, defined as
\begin{align}
        r\equiv \frac{\gamma_0^{\rm EMD}}{\gamma_0^{\rm RD}} = \sqrt{\frac{g_*(T_{\rm RH})}{g_*(T_{\rm kds})}} \left({\frac{g_*(T_{\rm dec})}{g_*(T_{\rm RH})}} \right)^{\frac{3}{8\alpha}}\left(\frac{T_{\rm kds}}{T_{\rm RH}}\right)^{2+n}\left(\frac{T_{\rm RH}}{T_{\rm dec}}\right)^{(4+n) - \frac{3}{2\alpha}}\,,
\end{align}
where $\gamma_{\rm el}=\gamma_0 T^{(4+n)}$. The three panels in Fig.~\ref{fig:free_stream} illustrate different scenarios in the presence of EMDE:
\begin{itemize}
    \item Left panel: \(\Gamma_\phi = \text{const.}\), resulting in \(T \sim a^{-3/8}\), \(H \sim T^4\), and partial decoupling due to \(p\)-wave scatterings (\(\gamma \sim T^6\)).
    \item Central panel: \(\Gamma_\phi \sim T\), leading to \(T \sim a^{-1/2}\), \(H \sim T^3\), and partial decoupling due to \(s\)-wave scatterings (\(\gamma \sim T^4\)).
    \item Right panel: \(\Gamma_\phi \sim T\), leading to \(T \sim a^{-1/2}\), \(H \sim T^3\), and full decoupling due to \(p\)-wave scatterings (\(\gamma \sim T^6\)).
\end{itemize}
We observe that in all three cases, the free-streaming horizon is smaller in the case of EMDE if $\gamma_0^{\rm EMD}\leq\gamma_0^{\rm RD}$, (i.e., $r\leq 1$). If the primary interaction governing the relic density of the DM also maintains it in kinetic equilibrium with the bath, one typically needs $r<1$ to match the observed relic density in both the early matter-dominated and the radiation-dominated cosmology. This requirement stems from the fact that the entropy injection during the EMDE phase dilutes the relic density, necessitating a comparatively lower interaction rate. However, in the presence of multiple interactions, one may fathom a situation where the processes involved in setting the relic density may not be the dominant interactions required for maintaining the kinetic equilibrium, such that $r>1$ is possible. In that case, $\lambda^{\rm EMD}_{\rm fsh}$ is smaller than $\lambda^{\rm RD}_{\rm fsh}$, only if $T_{\rm dec}/T_{\rm RH}$ is greater than a certain threshold value. 

In Sec.~\ref{DM_EMDE}, we will consider two simple models of dark matter where the DM scatters off the bath particles through $s$-wave and $p$-wave scattering, and compute the free-streaming horizon in terms of microscopic model parameters, in the presence of an EMDE with temperature-dependent entropy injection.

\section{Growth of matter perturbations:  halo formation}
\label{perturbations}

Now we focus on the particular scenario where $\Gamma_\phi\propto T$ during an EMDE and analyze the evolution of matter perturbations and formation of sub-earth halos. 

\subsection{Early matter domination: temperature dependent entropy injection}
\label{bkg_cosmo}

We consider a background cosmological model where the oscillating scalar field $\phi$, with the equation of state $\omega_\phi=0$, dominates the universe prior to the BBN and dissipates its energy into the radiation at a rate
\begin{eqnarray}
	\Gamma_{\phi}(a) = \Gamma_{0} T(a)\,,
\end{eqnarray}
where $\Gamma_0$ is a dimensionless constant. 
The time evolution of the energy densities of the meta-stable scalar field ($\rho_\phi$) and its decay products ($\rho_\gamma$) and the number density of the DM ($n_\chi$) are governed by the Boltzmann equations given below 
\begin{align}
\label{eqscalar}
	\frac{d\rho_{\phi}}{dt} + 3 H\rho_{\phi}  &= - \Gamma_{\phi}\rho_{\phi}\,, \\
\label{eqrad}
    \frac{d\rho_{\gamma}}{dt} + 4H\rho_{\gamma} &= \Gamma_{\phi}\rho_{\phi} + 2 \langle E \rangle \langle \sigma v \rangle_{\rm ann} \left(n_{\chi}^{2} - n_{\chi, {\rm eq}}^{2}\right), \\
\label{eqdm}
    \frac{d n_{\chi}}{dt} + 3Hn_{\chi} &= - \langle \sigma v \rangle_{\rm ann} \left(n_{\chi}^{2} - n_{\chi, {\rm eq}}^{2}\right),
\end{align}
where $\langle E \rangle = \sqrt{m_{\chi}^{2} + 9T^{2}}$ \cite{Erickcek:2015jza}, and the Hubble expansion rate $H^{2} = \left(\rho_{\phi} + \rho_{\gamma} + \rho_{\chi}\right)/3M_{\rm pl}^{2}$. 

The annihilation cross-section of the DM $\langle \sigma v\rangle_{\rm ann}$ can be calculated using Eq.~\eqref{eq:sigma_ann}, which depends on the specific microscopic model of the DM. We will focus on two simple dark matter models as case studies in Sec.~\ref{DM_EMDE}:
\begin{itemize}
    \item \underline{Model I ($s$-wave scattering):} Scalar dark matter $\phi_\chi$ annihilating into scalar radiation $\phi_\gamma$ via quartic $\phi_\chi^2\phi_\gamma^2$ interaction.
    \item \underline{Model II ($p$-wave scattering):} Fermionic dark matter $\psi_\chi$ annihilating into fermionic radiation $\psi_\gamma$, mediated by a heavy scalar $\varphi_M$, via $\bar\psi_\chi\psi_\gamma\varphi_M$ Yukawa interaction. 
\end{itemize}
In this section, we will present the results in a model-independent manner. We will explicitly indicate any instances where dependence on specific models arises.

At the early stages of reheating, the energy density of $\phi$ dominates as it coherently oscillates around the minima of the potential and subsequently dissipates its energy into radiation. The EMDE ends and the radiation domination starts at the reheating temperature ($T_{\rm RH}$) which is defined when $\rho_{\phi}(T_{\rm RH}) = \rho_{\gamma}(T_{\rm RH})$ \footnote{One can also define $T_{\rm RH}$ as the temperature when $\Gamma_{\phi}(T_{\rm RH}) = H(T_{\rm RH})$.}. 
The initial conditions (at $a=a_{\rm in}$) to solve the Boltzmann Eqs.~\eqref{eqscalar}, \eqref{eqrad}, and \eqref{eqdm} are given by
\begin{align}
    \rho_\phi(a_{\rm in})=3M_{\rm pl}^2H_{\rm in}^2\,, \quad \rho_{\gamma}(a_{\rm in})=0\,, \quad n_\chi(a_{\rm in})=0\,,
\end{align}
where $H_{\rm in}\equiv H(a_{\rm in})$ is a free parameter. The approximate solutions of the Eqs.~\eqref{eqscalar} and \eqref{eqrad} for $\Gamma_{\phi} \ll H$ are
\begin{align}
	\rho_{\phi}(a) \approx 3M_{\rm pl}^2H_{\rm in}^2\left(\frac{a}{a_{\rm in}}\right)^{-3}, ~
    \rho_{\gamma}(a) \approx \left[\left(\frac{\pi^2 g_{*}}{30}\right)^{-\frac{1}{4}}\frac{3M_{\rm pl}^2H_{\rm in}\Gamma_0}{2}\right]^{\frac{4}{3}}\left(\frac{a}{a_{\rm in}}\right)^{-4}\left[\left(\frac{a}{a_{\rm in}}\right)^{-\frac{3}{2}}-1\right]^{\frac{4}{3}}.
\label{sol_bkg}
\end{align}

We estimate the approximate energy density of $\gamma$ by considering only the term proportional to $\rho_{\phi}$ in the right-hand side of Eq.~\eqref{eqrad}. Eq.~\eqref{sol_bkg} shows that the temperature of the radiation ($T\propto \rho_\gamma^{1/4}$) quickly reaches a maximum $T_{\rm max}$ at $a_{\rm max}=(4/9)a_{\rm in}$, and then falls as $T\sim a^{-1/2}$ until the end of the reheating. The expressions for $T_{\rm max}$ and $T_{\rm RH}$ are given by
\begin{eqnarray}
	T_{\rm max} = \frac{1}{2}\left[\left(\frac{\pi^2 g_{*}}{30}\right)^{-1}3M_{\rm pl}^2H_{\rm in}\Gamma_0\right]^{1/3}\,, \quad T_{\rm RH} = \left(\frac{\pi^{2}g_{*}(T_{\rm RH})}{30}\right)^{-\frac{1}{2}}\frac{\sqrt{3}M_{\rm pl}\Gamma_0}{2}.
\end{eqnarray}

\begin{figure}[t!]
	\centering
	\includegraphics[width=0.6\textwidth]{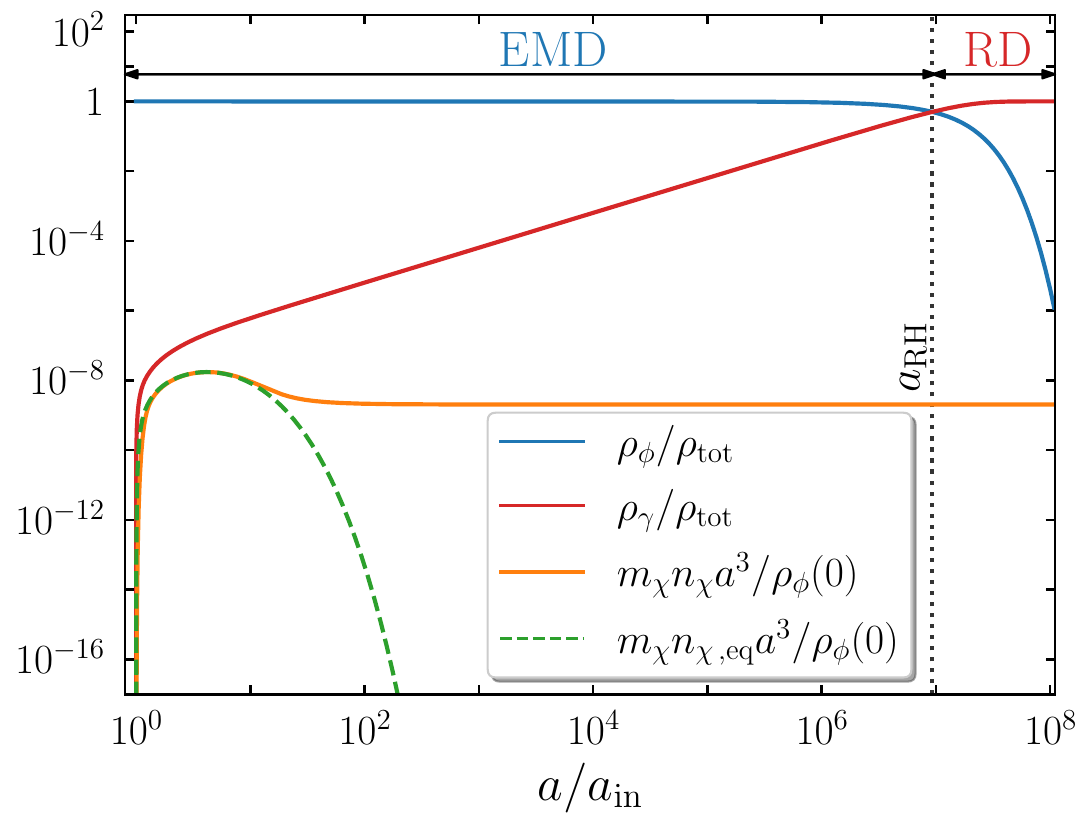}
	\caption{\sf\it Evolution of the background energy densities $\rho_\phi$ (blue) and $\rho_\gamma$ (red), normalized by the total energy density;  as well as the comoving number density of the DM $m_\chi n_\chi a^3/\rho_\phi(0)$ (orange). The equilibrium comoving number density of the DM is shown by the dashed green line. We set $m_{\chi} = 210 $ $GeV$, $ \langle \sigma v \rangle = 2.26 \times 10^{-14}\, GeV^{-2}$, $\lambda = 3.168 \times 10^{-4}$, and $T_{\rm RH} = 30$ $MeV$. The vertical dotted line marks the end of EMDE. 
  }
	\label{background}
\end{figure}

We consider that the DM remains non-relativistic at all times, thus we take $m_\chi \geq T_{\rm max}$. Despite initially having $n_\chi(0)=0$, the DM rapidly reaches equilibrium with the plasma owing to its substantial interactions with the bath particles and eventually undergoes chemical freeze-out around $T_{\rm fo}\sim m_\chi/20$. For illustration, we sketch the evolution of the background energy densities $\rho_\phi$ and $\rho_\gamma$ scaled by the total energy density in Fig.~\ref{background}, using the full numerical solutions of the set of Boltzman Eqs.~\eqref{eqscalar},~\eqref{eqrad}, and~\eqref{eqdm}. In Fig.~\ref{background}, we also show the evolution of the DM number density $n_\chi$ and compare it with the equilibrium number density $n_{\chi, {\rm eq}}$ for the Model I described in Sec.~\ref{scalarDM}.

\subsection{Evolution of matter perturbations}

The presence of EMDE has a significant impact on the density perturbations of the radiation and the dark matter. Especially, the matter perturbations show a linear growth for the modes that enter the horizon during the EMDE. We study the evolution of the density perturbations by linearly perturbing the continuity equation~\cite{Kodama:1984ziu, Gorbunov:2011zzc, Erickcek:2011us, Fan:2014zua, Erickcek:2015jza}
\begin{align}
    \nabla_\mu T^{\mu (i)}_\nu =Q^{(i)}_\nu, \quad i = \phi,\gamma, \chi\,.
\end{align}
Here, we assume the scalar $\phi$, radiation, and the DM behave as perfect fluids with energy momentum tensor $T^{(i)}_{\mu\nu}$. The energy exchange $Q^{(i)}_\nu$ occurs due to the dissipation of $\phi$, and annihilation of $\chi$. We work with the perturbed Friedmann–Lema\^{\i}tre–Robertson–Walker (FLRW) metric in the Newtonian gauge, 
\begin{eqnarray}
    \label{metic eq}
    ds^2 = -(1+2\Phi)dt^2 + a(t)^2(1-2\Psi)\delta_{ij}dx^idx^j\,,
\end{eqnarray} 
where $\Psi=\Phi$ in the absence of anisotropic stress. Equations governing the evolution of the fractional density perturbations ($\delta_i\equiv[\hat{\rho}_i(t,\vec{x})-\rho_i(t)]/\rho_i\equiv\delta\rho_i/\rho_i$), divergence of the velocity perturbations ($\theta_{i} \equiv a\vec{\nabla}\cdot \vec{v_i}$) for each fluid, and the metric perturbation ($\Phi$) are shown in the App.~\ref{sec:pert_eq}.

We are mainly interested in the modes (with wave number $k$) which are super-Hubble, i.e. $k\ll a H$ at $a_{\rm max}$, but subsequently enter the horizon during the EMDE epoch. Since the background energy density of the universe is dominated by $\phi$ at the onset of EMDE, we impose adiabatic initial conditions at $a_{\rm max}$ for the multi-fluid perturbations such that
\begin{equation}
    \label{adia_cond}
    \frac{\delta \rho_{\phi}}{\rho'_{\phi}}\Big|_{{a_{\rm max}}}=\frac{\delta \rho_{\gamma}}{\rho'_{\gamma}}\Big|_{{a_{\rm max}}}=\frac{\delta \rho_{\chi}}{\rho'_{\chi}}\Big|_{{a_{\rm max}}}\,,
\end{equation}
where prime denotes derivative with respect to the scale factor $a$. Therefore, the adiabatic initial conditions for the density perturbations of $k\ll aH$ modes are given by
\begin{align}
    \delta_\phi(a_{\rm max}) = - 2\Phi_0\,, ~ \delta_\gamma(a_{\rm max}) = \frac{2}{3}\delta_\phi(a_{\rm max})\,, ~ \delta_{\chi,{\rm eq}}(a_{\rm max}) = \frac{\delta_\gamma}{4}\left(\frac{3}{2}+\frac{m_\chi}{T}\right)\bigg|_{a_{\rm max}}\,, 
\end{align}
where $\Phi(a_{\rm max})=\Phi_0$. The velocity perturbations quickly decay outside of the Hubble radius, thus we set their initial values to zero at super-Hubble scales \cite{Fan:2014zua}. The initial condition for $\delta_\gamma$ is derived using the approximate scaling of the background solutions $\rho_\phi\sim a^{-3}$ and $\rho_\gamma\sim a^{-2}$ once the universe attains $T_{\rm max}$, see Eq.~\eqref{sol_bkg}. Since the DM was still in thermal equilibrium at $T_{\rm max}$, $\delta_{\chi}\simeq\delta_{\chi, {\rm eq}}$, which is obtained by varying the equilibrium number density $n_{\chi, {\rm eq}}$ with respect to temperature~\cite{Erickcek:2015jza}. 

\begin{figure}[t!]
     \centering
      \includegraphics[width=0.476\textwidth]{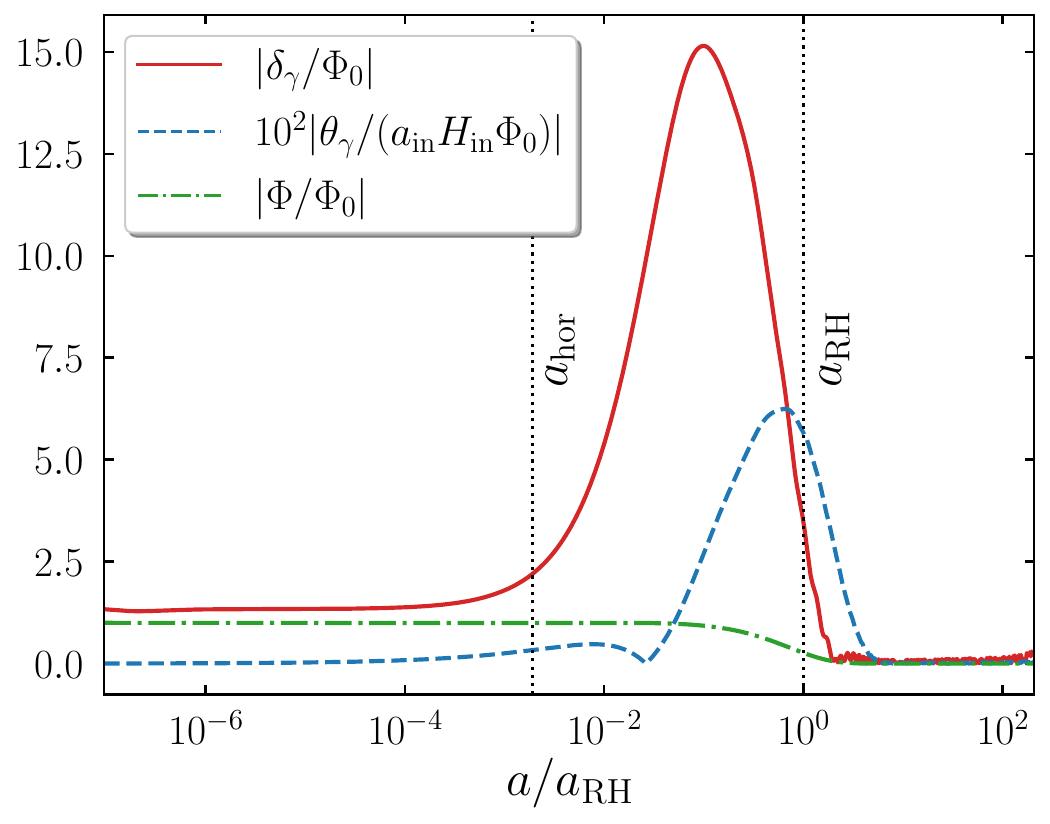}\hspace{5pt}
      \includegraphics[width=0.5\textwidth]{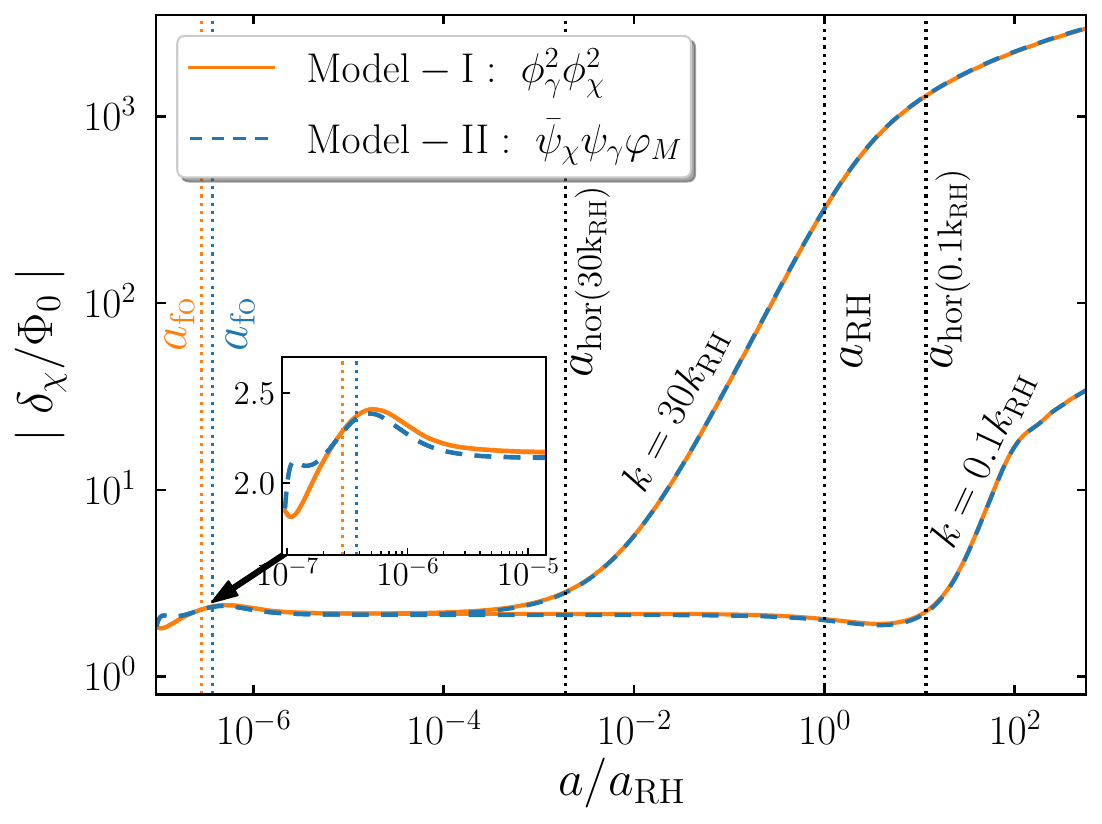}
     \caption{\sf\it Left: Evolution of $\delta_\gamma$ (red), $\theta_\gamma$ (blue dashed), and $\Phi$ (green dot-dashed) are shown as a function of $a/a_{\rm RH}$ for the mode $k = 30k_{\rm RH}$. The vertical line at $a=a_{\rm hor}$ signals the horizon entry of the mode. Right: The evolution of $\delta_\chi$ is presented for two modes: one entering during an EMDE ($k=30 k_{\rm RH}$) and the other during RD ($k = 0.1 k_{\rm RH}$). The results are shown for both Model I (orange) and Model II (blue dashed), fixing $T_{\rm RH} = 10$ MeV, $H_{\rm in} = 5.786\times 10^{-12}$ GeV, $m_{\chi} = 95$ GeV, and $\lambda = 4.912\times10^{-5}$ (Model I), while $y = 7.85\times10^{-3}$, $M = 175$ GeV (Model II). The inset magnifies the evolution of $\delta_\chi$ near the freeze-out.}
        \label{fig:perturb}
\end{figure}

In Fig.~\ref{fig:perturb}, we present the numerical solutions of the perturbation Eqs.~\eqref{perturbeq} to \eqref{phi}, in Fourier space as a function of $a/a_{\rm RH}$ using a benchmark parameter set. In the left panel, the evolution of $\Phi$, $\delta_\gamma$, and $\theta_\gamma$ are shown for a mode $k=30\;k_{\rm RH}$, where $k_{\rm RH}=a_{\rm RH}H(a_{\rm RH})$ represents the wave number of the mode that enters the horizon at $T_{\rm RH}$. The $k=30\;k_{\rm RH}$ mode enters the horizon at $a\simeq 1.9\times10^{-3}a_{\rm RH}$ while the reheating concludes at $a_{\rm RH}/a_{\rm in}=1.7\times 10^{7}$. We observe that $\Phi$ remains nearly constant during the EMDE, while $\delta_\gamma$ receives an initial kick during the horizon crossing and starts to grow due to the source term arising from the dissipation of $\phi$, see Eq.~\eqref{deltar}. However, this growth in $\delta_\gamma$ does not sustain till the end of EMDE due to a back-reaction from the growing spatial dispersion (increasing $\theta_{\gamma}$) of the radiation fluid. All in all, $\delta_{\gamma}$ reaches a peak and then decays before the reheating ends, and oscillates with a suppressed amplitude once the radiation domination begins. 

The approximate solutions for the evolution of $\theta_\phi$, $\theta_\chi$ and $\delta_\phi$ during the EMDE is obtained in the limit $\Gamma_\phi \ll H$ as
\begin{eqnarray}
        \label{thetascalar2}
        \theta_{\phi}\simeq \theta_{\chi}\simeq A(k){a_{\rm in}}{H_{\rm in}}\Phi_0 \left(\frac{a}{a_{\rm in}}\right)^{\frac{1}{2}}\,, \,~\, \delta_{\phi}\simeq -2\Phi_0-A(k)\Phi_0 \frac{a}{a_{\rm in}}\,, \,~\, A(k) \equiv \frac{2k^2}{{3a_{\rm in}^2}{H_{\rm in}^2}}\,.
    \end{eqnarray}
The density perturbation of the DM ($\delta_\chi$) follows the $\delta_{\chi, {\rm eq}}$ as long as the DM remains in chemical equilibrium. However, after the freeze-out ($a>a_{\rm fo}$), $\delta_\chi$ decreases to $-2\Phi_0$ and remains constant until the corresponding mode enters the horizon. This is consistent with the adiabatic initial condition in Eq.~\eqref{adia_cond}, since $\rho_{\phi,\chi}\sim a^{-3}$ after freeze-out. After the horizon entry of the mode ($a>a_{\rm hor}$), $\delta_\chi$ grows linearly until reheating, followed by a logarithmic scaling during radiation domination, and resumes linear growth after matter-radiation equality for $a>a_{\rm eq}$. The leading scaling behavior of $\delta_\chi$ with the scale factor at different epochs are summarized below
\begin{align}
   \label{deltachi}
    \delta_\chi \sim \begin{cases} \delta_{\chi, {\rm eq}}, & a \leq a_{\rm fo}\,,\\
    -2\Phi_0, & a_{\rm fo}\leq a\leq a_{\rm hor}\,,\\
    -A(k)\Phi_0 \dfrac{a}{a_{\rm in}}, & a_{\rm hor} \leq a\leq a_{\rm RH}\,,\\
    -A(k)\Phi_0 \dfrac{a_{\rm RH}}{a_{\rm in}}\ln \left(\dfrac{a}{a_{\rm RH}}\right), & a_{\rm RH}\leq a \leq a_{\rm eq}\,,\\
    -A(k)\Phi_0 \dfrac{a_{\rm RH}}{a_{\rm in}}\ln \left(\dfrac{a_{\rm eq}}{a_{\rm RH}}\right)\dfrac{a}{a_{\rm eq}}, & a \geq a_{\rm eq}\,.\\
    \end{cases}
\end{align}
In the right panel of Fig.~\ref{fig:perturb} we present a full numerical solution of the evolution of $\delta_\chi$ for the modes $k=30 k_{\rm RH}$ and $k= 0.1 k_{\rm RH}$, where the former enters the horizon before reheating and the latter during the radiation domination. The linear growth of the mode $k=30 k_{\rm RH}$ from its horizon entry until the end of reheating amplifies $\delta_\chi$ compared to the mode that enters the horizon during radiation domination. Clearly, the evolution of $\delta_\chi$ for both the scalar (Model I) and fermionic (Model II) dark matter models coincides once the dark matter decouples chemically, since the model-dependent term in the Eq.~\eqref{deltadm}, i.e., $\Sigma^{\delta_\chi}_{\rm ann}$ becomes negligible after freeze-out.

\subsection{Matter power spectrum}

Formation of the small-scale structures due to the linear growth of matter perturbations for the modes that become sub-Hubble during the EMDE start well after the universe went through the matter-radiation equality. The effect of the EMDE enters into the matter power spectrum through a mode-dependent rescaling of the transfer function $T(k)$ as $T(k) \to R(k) T(k)$, where $R(k)$ accounts for the boost in the matter power spectrum due to the linear growth of perturbations during the EMDE.

To compute $R(k)$, we first determine the variation of the DM perturbations with the wave number of the modes using a smoothly interpolating empirical function as \cite{Erickcek:2011us}
\begin{align}  
    \label{fit function}
    \delta_{\chi} = \frac{10}{9}C_1(k/k_{\rm RH})\Phi_0\ln\left[C_2(k/k_{\rm RH})\frac{a}{a_{\rm hor}}\right]\,.
\end{align}
Here $C_1(k)$ and $C_2(k)$ are two empirical functions, as shown in Eq.~\eqref{fit function_2}, with unknown parameters $\alpha_i$ which are obtained by fitting against the full numerical solutions of $\delta_\chi(k)$ at a fixed time well after the reheating, see App.~\ref{sec:pert_eq} for more details. Note that we require somewhat different empirical functions $C_1(k)$ and $C_2(k)$, compared to \cite{Erickcek:2011us} due to the different background evolution near reheating. In the standard radiation dominated universe, $C_1 = 9.11$ and $C_2 = 0.594$, as given by \cite{Hu:1995en}. 

To take into account the non-zero baryon fraction, we obtain the final expression of $R(k)$ for $k>0.05\,k_{\rm RH}$ by matching the growing and decaying solutions of Meszaros equation for $\delta_{\chi}$  with Eq.~\eqref{fit function} as \cite{Bertschinger:2006nq, Erickcek:2011us, Hu:1995en}.
\begin{eqnarray}
    \label{Rk}
    R\left(k\right) = \frac{C_1\left(\frac{k}{k_{\rm RH}}\right)\log\left[\left(\frac{4}{e^3}\right)^{\frac{f_2}{f_1}}C_2\left(\frac{k}{k_{\rm RH}}\right)\frac{a_{\rm eq}}{a_{\rm hor}}\right]}{9.11\log\left[\left(\frac{4}{e^3}\right)^{\frac{f_2}{f_1}}0.594\frac{\sqrt{2}k}{k_{\rm eq}}\right]}\,,~ \begin{cases}  
    & f_b = \frac{\Omega_{\rm b}}{\Omega_{\chi}+\Omega_{\rm b}}\,, \\
    & f_1 = 1-0.568 f_b+0.094f_b^2\,,\\
    & f_2 = 1-1.156f_b+0.149f_b^2-0.074f_b^3\,.
    \end{cases}
\end{eqnarray}
Here $\Omega_\chi$ and $\Omega_b$ denote the relic density of the DM and the baryons, respectively. The scale factor at the horizon entry $a_{\rm hor}(k)$ for modes with different scales are given by
\begin{eqnarray}
   \label{aeqnyahor}
    \dfrac{a_{\rm eq}}{a_{\rm hor}} = \begin{cases}0.7\left(\dfrac{k}{k_{\rm RH}}\right)^2\dfrac{T_{\rm RH}}{T_{\rm eq}}\left(\dfrac{g_{*S}\left(T_{\rm RH}\right)}{g_{*S}\left(T_{\rm eq}\right)}\right)^{(1/3)}, & k\geq k_{\rm RH}\,,\\\\
    0.8\dfrac{\sqrt{2}k}{k_{\rm eq}}, & k_{\rm eq}\leq k < k_{\rm RH}\,,\\
    \end{cases}
\end{eqnarray}
where $k_{\rm RH}$ is given by
\begin{align}
    \label{krhbykeq}
     \dfrac{k_{\rm RH}}{k_{\rm eq}} =\dfrac{7.11\times10^9}{1+z_{\rm eq}}\left(\dfrac{T_{\rm eq}}{1\; \rm eV } \right) \left(\dfrac{T_{\rm RH}}{1\; \rm MeV}\right)\left(\dfrac{10.75}{g_{*S}\left(T_{\rm RH}\right)}\right)^{1/3}\left(\dfrac{g_{*}\left(T_{\rm RH}\right)}{10.75}\right)^{1/2}.
\end{align}
\begin{figure}[t!]
     \centering
      \includegraphics[width=0.65\textwidth]{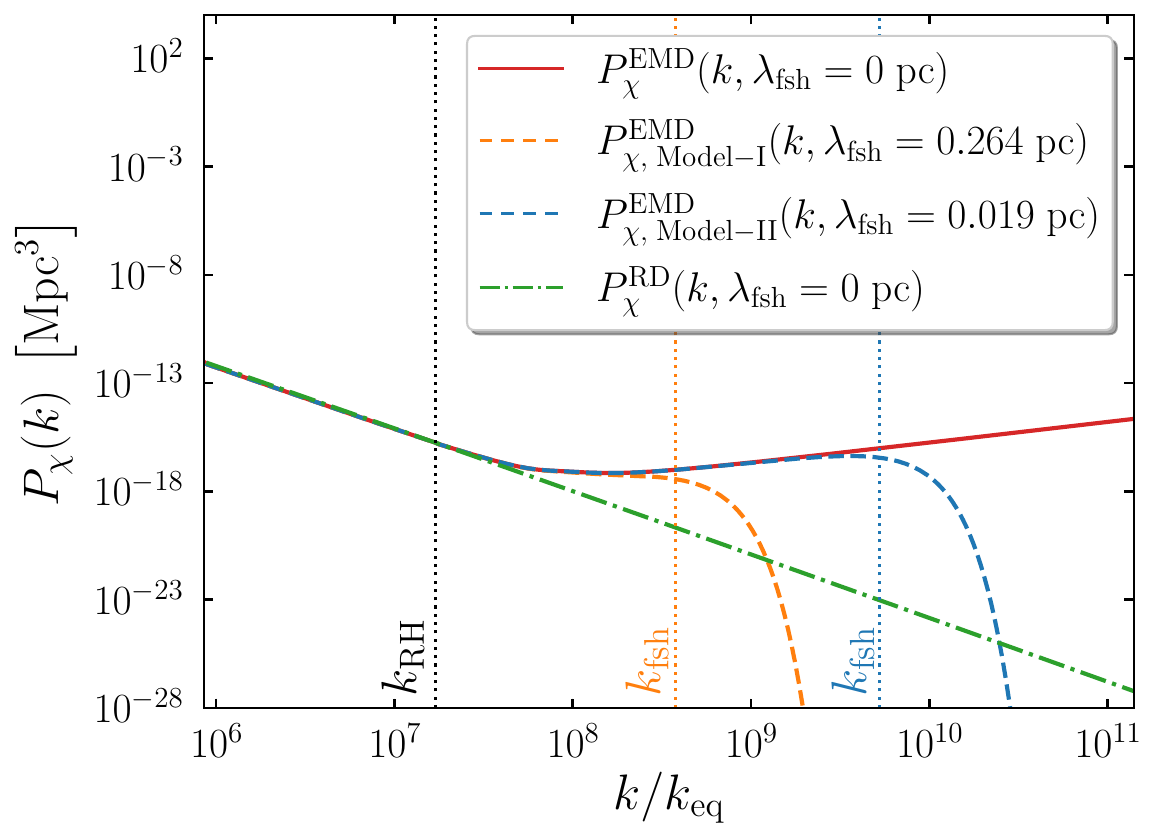}
     \caption{\sf\it 
     The matter power spectrum $P_\chi(k)$ for different values of $\lambda_{\rm fsh}$, evaluated at $z=50$, in case of an EMDE with a temperature-dependent entropy injection, where $T_{\rm RH} = 10$ MeV, and $k_{\rm RH}=1.7\times10^7\;k_{\rm eq}$. We use the same set of benchmark points as mentioned in Fig.~\ref{fig:perturb}. The green dot-dashed line represents $P_\chi(k)$ for a purely radiation-dominated universe with $\lambda_{\rm fsh}=0$. The coloured vertical lines correspond to the modes for which $k\sim \lambda_{\rm fsh}^{-1}$, and have the values $k_{\rm fsh}=3.8\times10^8\;k_{\rm eq}$ and $5.3\times10^9\;k_{\rm eq}$, for Model I and Model II, respectively. The matter power spectrum $P_\chi(k)$ shows different scaling behavior on two sides of $k_{\rm RH}$ following Eq.~\eqref{power standard}.}
        \label{fig:power spectrum}
\end{figure}
The green dot-dashed curve in Fig.~\ref{fig:power spectrum} shows the variation of the power spectrum $P_\chi(k,z)$ with $k/k_{\rm eq}$ in the standard radiation dominated universe, evaluated at the redshift $z=50$ and keeping $\lambda_{\rm fsh}=0$. The primordial curvature power spectrum, $P_\mathcal{R}(k)\propto |\Phi_0|^2$, is defined as  
\begin{eqnarray}
\label{R}
    \Delta_{\mathcal{R}} = \frac{k^3}{2\pi^2}P_{\mathcal{R}} = A_s\left(\frac{k}{k_0}\right)^{\left(n_s-1\right)},
\end{eqnarray}
where $A_s = 2.1\times10^{-9}$, $n_s = 0.965 $ and $k_0 = 0.05$ Mpc$^{-1}$, are taken from the Planck data \cite{Planck:2018vyg}. Following \cite{Erickcek:2011us}, the power spectrum is calculated using CAMB \cite{Lewis:2007kz}\footnote{\url{https://camb.info/}} for $k/k_{\rm eq} \leq 8.2 \times 10^5$, while we use the transfer function given in \cite{Eisenstein_1998}\footnote{\url{http://background.uchicago.edu/~whu/transfer/transferpage.html}} for $k/k_{\rm eq} > 8.2 \times 10^5$. We used the Planck data for the present-day Hubble parameter $H_0 = 67.4$ km s$^{-1}$ Mpc$^{-1}$, relic density of the baryons $\Omega^0_{\rm b}h^2 = 0.0224$, the dark matter relic density $\Omega^0_{\rm DM}h^2 = 0.120$, and $k_{\rm eq} = 0.01$ Mpc$^{-1}$  \cite{Planck:2018vyg}.
Finally, we multiply the power spectrum for the redshift $500\leq z\leq 3$, by a scale-dependent growth function evaluated w.r.t. $z = 50$, as defined in \cite{Erickcek:2011us}.

The solid red curve in Fig.~\ref{fig:power spectrum} presents the matter power spectrum in the presence of EMDE with energy-dependent dissipation. The leading momentum dependence of $P_\chi(k) \propto |\delta_\chi|^2$ for the modes entering the horizon before and after reheating is given by
\begin{eqnarray}
    \label{power standard}
    P_{\chi}\left(k, a\gg a_{\rm eq}\right) \propto \begin{cases}
    k^4 P_{\mathcal{R}}
    \propto k^{n_s}, & k>k_{\rm RH},\\\\
    \left[\ln\left(\sqrt{2}\frac{k}{k_{\rm eq}}\right)\right]^2k^{\left(n_s-4\right)}, & k<k_{\rm RH}.
    \end{cases}
\end{eqnarray}

We assume the free-streaming horizon of the DM is zero for the red curve. However, in practice, the DM begins to free-stream after decoupling  (both kinetically and chemically) from the thermal bath, resulting in a non-zero free-streaming horizon. To account for the washout of structures on scales smaller than the free-streaming horizon $\lambda_{\rm fsh}$ of the DM, the power spectrum is modulated with a Gaussian cut-off as $P_\chi(k)\rightarrow \exp\left(-k^2/k^2_{\rm fsh}\right)P_\chi(k)$, where $k_{\rm fsh}=\lambda_{\rm fsh}^{-1}$. The orange and blue dashed curves in Fig.~\ref{fig:power spectrum} illustrate the exponential decline of the power spectrum for modes with $k > k_{\rm fsh}$, corresponding to two values of $\lambda_{\rm fsh}$ that represent specific benchmark points for the scalar (Model I) and fermionic (Model II) DM models, respectively. Thus, the modes receiving enhancement of the power spectrum due to the linear growth of matter perturbations during EMDE is bounded by a lower cut-off $k_{\rm RH}$, set by the reheating temperature and an upper cut-off $k_{\rm fsh}$, set by the free-streaming horizon of the DM.

\subsection{Formation of sub-earth halos}

The enhancement of power spectrum for the modes $k_{\rm RH}<k<k_{\rm fsh}$ dictates the range of mass scales for the DM halos having an enhanced population compared to the standard radiation-dominated universe. Two characteristic mass scales, setting the limits on the masses contained in the halos with enhanced populations are
\begin{itemize}
    \item $M_{\rm RH}$, which determines the halos with maximum mass, and of size $R_{\rm RH}\sim k_{\rm RH}^{-1}$,
    \item $M_{\rm fsh}$, representing the minimum mass halos of size $R_{\rm fsh}\sim k_{\rm fsh}^{-1}$. 
\end{itemize}

The rms density perturbation in a spherical DM halo of radius $R$, and average mass $M = (4\pi/3)R^3\rho^0_{\chi}$, where $\rho^0_{\chi}$ is the present-day dark matter density, is given by 
\begin{eqnarray}
    \label{sigma}
    \sigma^2(M,z) = \int_0^{\infty}\frac{\mathrm{d}^3k}{\left(2\pi\right)^3}P(k,z)F^2(kR).
\end{eqnarray}
We use a filter function $F(kR)$, following \cite{Erickcek:2011us}, where a top-hat window function is convoluted with a Gaussian window function such that the modes larger than $R^{-1}$ do not contribute to the integral in Eq.~\eqref{sigma}. 

\begin{figure}[t!]
     \centering
      \includegraphics[width=0.475\textwidth]{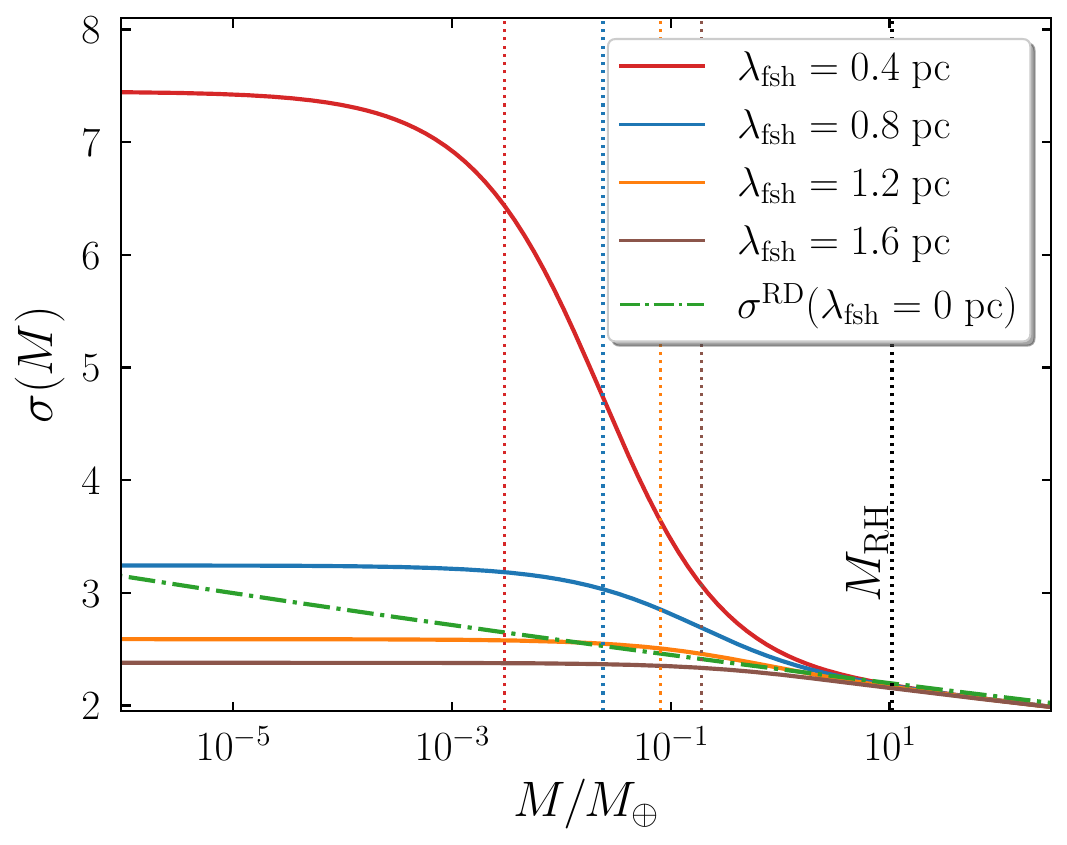}\hspace{5pt}
      \includegraphics[width=0.5\textwidth]{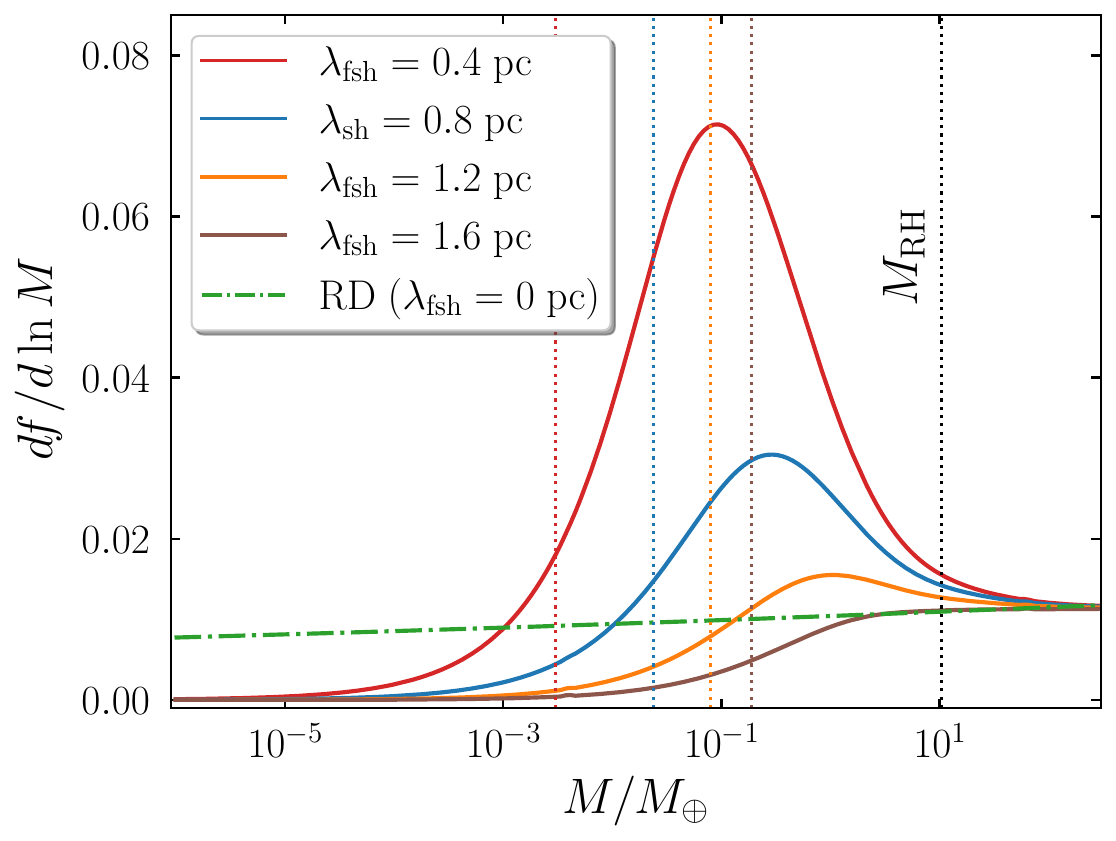}
     \caption{\sf\it  $\sigma(M)$ (left) and differential mass fraction (right) at $z = 10$ are shown as a function of dark matter halo mass in units of earth mass($M_\oplus$), for different values of $\lambda_{\rm fsh}$. We set $T_{\rm RH} = 10$ MeV which corresponds to $M_{\rm RH} = 10.5\, M_{\oplus}$. The vertical dotted contours denote the value of $M_{\rm fsh}$ associated with the respective free-streaming horizons. The green dot-dashed curves represent the scenario with pure radiation domination and $\lambda_{\rm fsh}=0$.}
        \label{fig:sigma}
\end{figure}

The left panel of Fig.~\ref{fig:sigma} presents the variation of $\sigma(M)$ with the halo mass $M$ normalized to the mass of the earth ($M_\oplus$) for different values of the free-streaming horizon, at a redshift $z=10$, and $T_{\rm RH} =10$ MeV. Clearly, for $M>M_{\rm RH}$, $\sigma(M)$ follows the rms density that would be found in a standard radiation-dominated universe with $\lambda_{\rm fsh}=0$. However, for $M_{\rm fsh}<M<M_{\rm RH}$, a power law growth $\sigma\left(M\right)\propto M^{-\left(n_s+3\right)/6}$ is observed, which becomes constant at scales below $M_{\rm fsh}$. 

The differential fraction of DM mass ($df_{\rm halo}$) contained in a halo within mass range $M$ and $M+\mathrm{d}M$ is given \`a la Press-Schechter as \cite{1974ApJ...187..425P}
\begin{eqnarray}
    \label{mass function}
    \frac{\mathrm{d}f_{\rm halo}}{\mathrm{d}\ln(M)} = \sqrt{\frac{2}{\pi}}\left|\frac{\mathrm{d}\ln\sigma}{\mathrm{d}\ln(M)}\right|\frac{\delta_c}{\sigma\left(M,z\right)}\exp\left(-\frac{\delta_c^2}{2\sigma^2\left(M,z\right)}\right),
\end{eqnarray}
where $\delta_c\simeq 1.686$ is the critical value of linear overdensity for $z\geq 2$. In the right panel of Fig.~\ref{fig:sigma} we show that the increase in the differential fraction significantly depends on the free-streaming horizon of the DM. In fact, we observe that if $\lambda_{\rm fsh}$ is below a threshold value $\lambda_{\rm fsh}^{c}$, the structures forming at scales smaller than $M_{\rm RH}$ in the presence of EMDE are guaranteed to show a boost in the differential fraction compared to the radiation dominated universe. The threshold $\lambda_{\rm fsh}^{c}\simeq 1.6$ pc primarily depends on $T_{\rm RH}$. Since we determine the threshold $\lambda_{\rm fsh}^{c}$ by comparing with a scenario where $\lambda_{\rm fsh}=0$ in the purely radiation-dominated universe, any $\lambda_{\rm fsh}<\lambda_{\rm fsh}^{c}$ in the case of EMDE ensures a boost in the structure formation.

Enhanced small-scale structures could amplify DM annihilation signals within sub-earth halos, providing a probe for a variety of DM models as well as non-standard cosmological epochs like EMDE. The dynamics of DM decoupling during EMDE determine the free-streaming horizon, which, when compared to the threshold horizon $\lambda_{\rm fsh}^{c}$, can help in identifying a viable range of microscopic model parameters compatible with enhanced structures, we defer further discussion to Sec.~\ref{DM_EMDE}. 
\begin{figure}[t!]
     \centering
      \includegraphics[width=0.65\textwidth]{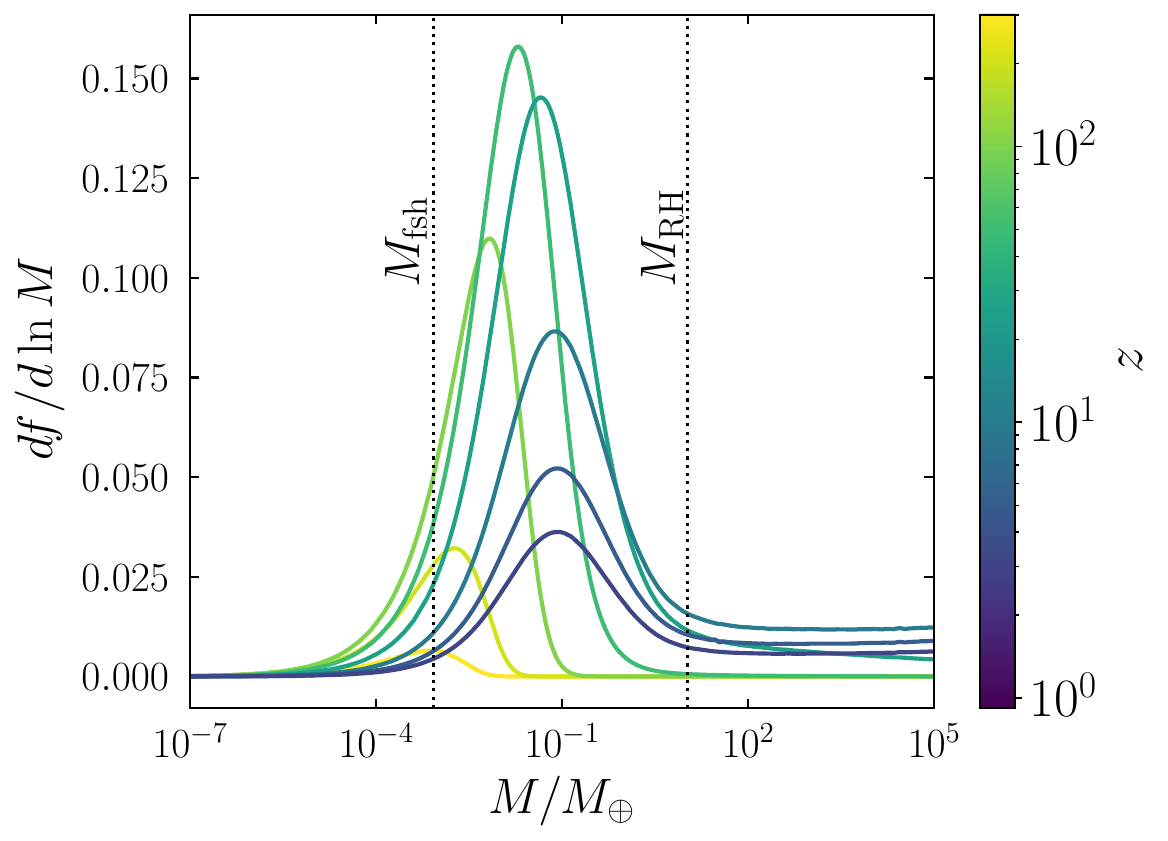}
     \caption{\sf\it Differential halo mass fraction is presented in the mass range $\in(10^{-7}-10^{5})M_\oplus$  for redshift ranging from $z=300$ to $z=3$, for $\lambda_{\rm fsh} = 0.264$ pc. This free-streaming horizon corresponds to $T_{\rm RH} = 10$ MeV, $m_{\chi} = 95$ GeV, $\lambda = 4.912\times10^{-5}$ for Model I.} 
        \label{fig:mass_funct}
\end{figure}

Finally, in Fig.~\ref{fig:mass_funct}, we illustrate how the differential mass fraction varies with halo mass for different redshifts, using $T_{\rm RH}=10$ MeV and $\lambda_{\rm fsh}=0.264$ pc (corresponding to $M_{\rm fsh}=8.5 \times 10^{-4} M_\oplus$), as an example. The first halos begin to form around $z=300$, and the peak of the differential fraction $M_{\rm peak}$ increases until around $z=50$, when $\sigma(M_{\rm peak},z) \sim \delta_c$. Beyond this point, the earlier forming halos start to merge, leading to a decrease in the differential fraction. Ultimately, $M_{\rm peak}$ saturates at around $M_{\rm peak} \simeq 0.1\, M_\oplus$ for $z < 10$, while the abundance of structures with $M \gg M_{\rm RH}$ remains the same as in the radiation-dominated universe.

\section{Case studies: scalar and fermionic dark matter}
\label{DM_EMDE}

To complement our model-independent discussion so far, in this section we present some underlying particle physics models to describe the interaction between the thermal plasma and the non-relativistic dark matter. In particular, we consider two simple examples exhibiting $s$-wave and $p$-wave elastic scatterings, 
for illustrative purposes. More detailed microscopic models including other types of interactions~\cite{Bringmann:2016ilk}, can also result in $s$- and $p$-wave elastic scattering rates. Similar analyses can be applied to these models as well. We identify regions in the microscopic parameter space of these models where the kinetic decoupling occurs during the EMDE (with $\Gamma_\phi\sim T$) and the sub-earth halo formation amplifies. Additionally, we highlight the key differences in the DM model space that arise from scenarios where the entropy injection rate during the EMDE is constant compared to where it is temperature-dependent. 

\subsection{Model I: s-wave scattered scalar dark matter}
\label{scalarDM}

The first model we consider has a scalar dark matter ($\phi_\chi$), interacting with scalar bath particles ($\phi_\gamma$) via a quartic vertex $(\lambda/4) \phi_\chi^2\phi_\gamma^2$. In general, $\phi_\gamma$ may represent either the Standard Model Higgs boson or any BSM scalar in the thermal bath. However, in the following, we assume that the mass of $\phi_\gamma$ is much smaller than all the relevant scales. This leads to an $s$-wave scattering $\phi_\chi\phi_\gamma\to\phi_\chi\phi_\gamma$, with momentum transfer rate
\begin{eqnarray}
	\gamma_{\rm el}(T) = \frac{\lambda^{2}\pi}{180}m_{\chi}\left(\frac{T}{m_{\chi}}\right)^{4}\,,
\end{eqnarray}
where $m_\chi\in(20-1000)$ GeV is the typical mass for a WIMP DM.

The thermally averaged cross-section for the annihilation process $\phi_\chi\phi_\chi \to \phi_\gamma\phi_\gamma$, which sets the relic abundance is computed using Eq.~\eqref{Eq:th_avg_crss} as
\begin{eqnarray}
	\left\langle \sigma v \right \rangle_{\rm ann}
	& \simeq & \frac{\lambda^{2}}{32\pi m_{\chi}^{2}} \left(1 - 3 \frac{T}{m_{\chi}}\right).
\end{eqnarray}
\begin{figure}[t!]
     \centering
      \includegraphics[width=0.8\textwidth]{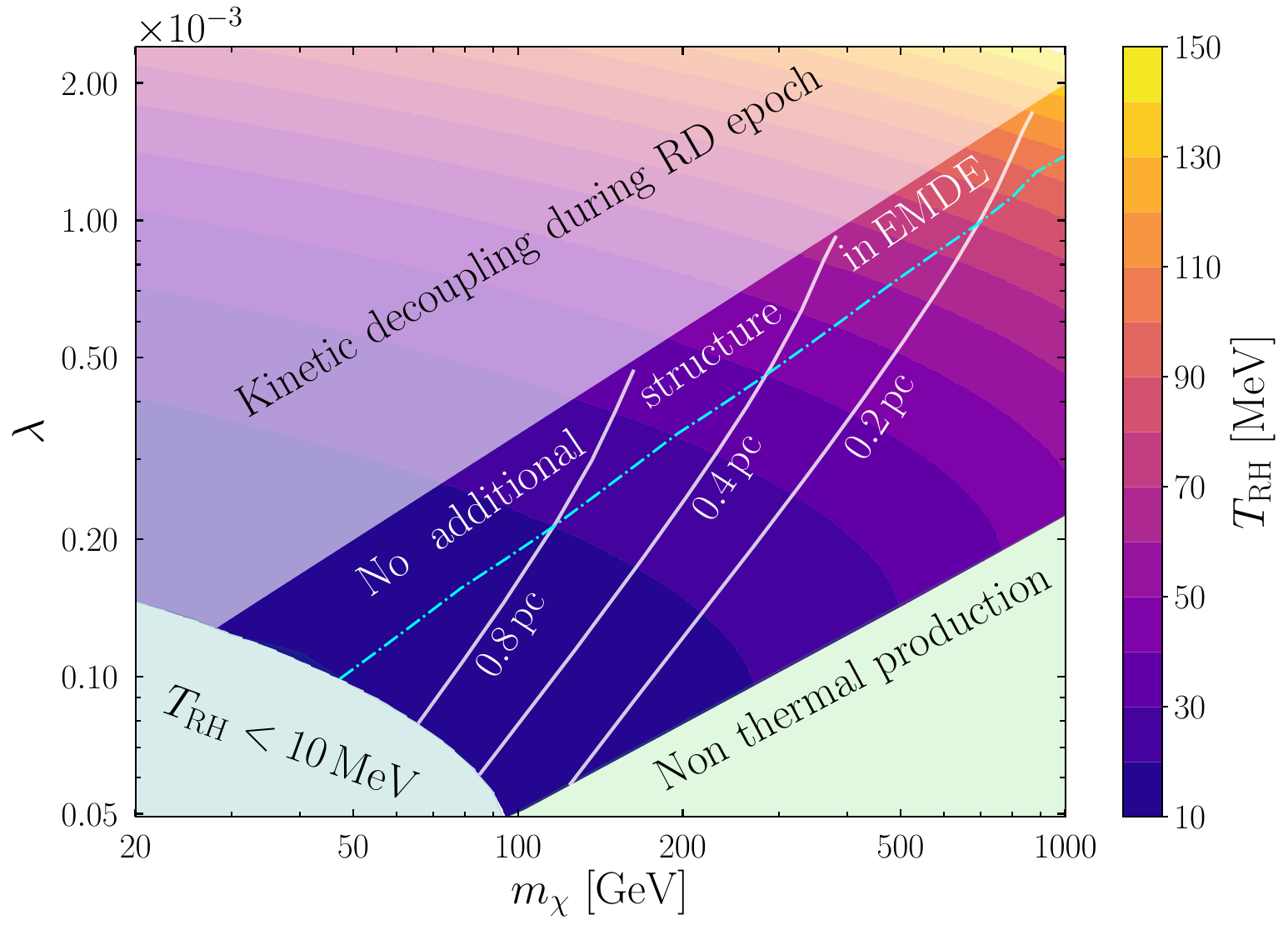}
     \caption{\sf\it The allowed parameter space satisfying the present-day relic density of the DM is shown with distinct values of $T_{\rm RH}$ presented by coloured bands. The DM shows partial kinetic decoupling during the EMDE in the dark-shaded region, while in the light-shaded region, DM kinetically decouples after reheating. In the lime coloured region below the colour bands, DM produces non-thermally from the bath, while the light cyan region, where $T_{\rm RH}< 10$ MeV, is excluded to avoid bounds from the BBN. The isocontours of the free-streaming horizon are shown by the white curves. Amplification of small-scale structures compared to the standard radiation-dominated cosmology occurs to the right of the cyan dot-dashed contour, which represents the cut-off value of the free-streaming horizon $\lambda^c_{\rm fsh}$.}
        \label{figure_point}
\end{figure}
In Fig.~\ref{figure_point}, we categorized the allowed regions in the $\lambda-m_\chi$ parameter space satisfying the following constraints:
\begin{itemize}
    \item EMDE concludes at $T_{\rm RH}\geq 10$ MeV to evade strong bounds from the BBN~\cite{Sarkar:1995dd, Hannestad:2004px}, isocontours of $T_{\rm RH}$ are shown in colour.
    \item The DM was in thermal contact at the onset of the EMDE, and freezes-out prior to the end of reheating, the observed present-day relic density of the DM $\Omega^{0}_{\rm DM}h^{2}=0.12$~\cite{Planck:2018vyg} is reproduced in the entire coloured region.
    \item In the dark-coloured area, partial kinetic decoupling of the DM occurs, however, the sub-earth halo population amplifies below the cyan dot-dashed contour.
    \item The isocontours of the free-streaming horizon $\lambda_{\rm fsh}$ (white curves), intersect the cyan contour at the threshold value $\lambda^c_{\rm fsh}$ for fixed $T_{\rm RH}$.
\end{itemize}
A few comments regarding the impact of the temperature-dependent entropy injection during the EMDE are in order. First of all, the annihilation cross-section required to satisfy the relic density in this cosmological background is smaller than that required in a fully radiation-dominated universe, due to the dilution of the relic by the entropy injection. In fact, in the parameter space of Fig.~\ref{figure_point} that shows freeze-out production of the DM, $\langle\sigma v\rangle_{\rm ann}$ ranges from $10^{-11}-10^{-15}$ GeV$^{-2}$ depending on $m_\chi$, which is at least two orders of magnitude smaller than what would be required in the absence of EMDE. This reduction in the annihilation cross-section helps evade the strong bounds coming from the direct detection experiments. On the other hand, we emphasize that the required cross-section in this scenario is in principle larger than that in an EMDE with a constant rate of entropy injection ($\Gamma_\phi \sim {\rm const.}$). To give an idea, we estimate the ratio of $\langle \sigma v\rangle_{\rm ann}$, required to produce the same comoving number density of the DM after reheating in EMDEs with constant and temperature-dependent entropy injection rates as follows

\begin{equation}\label{eq: cross-section compare final}
	\frac{\langle \sigma v\rangle^{\Gamma_\phi\sim T}_{\rm ann}}{\langle \sigma v\rangle^{\Gamma_\phi\sim {\rm const.}}_{\rm ann}} \sim  \frac{m_{\chi}}{10\, {\rm GeV}}\frac{10\, {\rm MeV}}{T_{\rm RH}}\frac{10^3}{x^c_{\rm fo}}\left(\frac{x^T_{\rm fo}}{x^c_{\rm fo}}\right)^3\,.
\end{equation}

Here $x^T_{\rm fo}$ and $x^c_{\rm fo}$ denote the ratio $m_\chi/T_{\rm fo}$ in the two scenarios with $\Gamma_\phi\sim T$ and $\Gamma_\phi\sim {\rm const.}$, respectively. Assuming that the freeze-out occurs around $x_{\rm fo}\sim 20$ in both scenarios and neglecting the $T/m_\chi$ dependence in $\langle\sigma v\rangle_{\rm ann}$, we find that the annihilation cross-section required for the same amount of relic is approximately 50 times larger in the $\Gamma_\phi\sim T$ case compared to the $\Gamma_\phi\sim {\rm const.}$ case. However, we also point out that the present-day relic density of the DM, $\Omega^{0}_{\rm DM}h^{2}=0.12$~\cite{Planck:2018vyg}, cannot be satisfied by the quartic interaction during an EMDE with $\Gamma_\phi\sim {\rm const.}$ if $T_{\rm RH}\ll 10$ GeV~\cite{Giudice:2000ex}. Therefore, the temperature-dependent entropy injection accommodates a novel region of the parameter space where the relic density can be satisfied even with low reheating temperature $T_{\rm RH}\sim 100$ MeV and reduced annihilation cross-section compared to the radiation-dominated universe\footnote{The constraints on $T_{\rm RH}$ derived in \cite{Choi:2017ncz}, based on fixed values for the WIMP mass, its annihilation cross-section, and the microhalo mass, cannot be directly applied to the present scenario due to the difference in the underlying assumptions, thus, we leave a detailed analysis of such constraints for future work.}. 

The above discussion indicates the crucial impact of the background cosmological epoch on the production of DM. Moreover, in a large range of $m_\chi$, the DM with $s$-wave scattering shows partial kinetic decoupling during the EMDE. This is also unique for $\Gamma_\phi\sim T$ scenario, as $s$-wave partial amplitudes are not sufficient to kinetically decouple the DM in $\Gamma_\phi\sim {\rm const.}$ scenario, see Table~\ref{table:QD_poss}. Thus, a broad region of the parameter space exhibits the correct freeze-out relic density with a low reheating temperature, leading to a reduced DM free-streaming horizon below the cut-off $\lambda^c_{\rm fsh}$. The resulting enhancement in small-scale structure formation at sub-earth mass scales offers new avenues to probe this scenario through DM annihilation signatures, provided the boost in the density of sub-halos can compensate for the reduced annihilation cross-section of the DM. However, alternate probes based on gravitational interaction of the DM such as pulsar timing arrays~\cite{Siegel:2007fz, Baghram:2011is, Choi:2017ncz, Dror:2019twh}, microlensing~\cite{Blinov:2021axd, Bechtol:2022koa}, as well as the sub-halo DM interactions with cosmic rays~\cite{Bramante:2021dyx}, and sub-halo collisions with neutron stars could provide additional handles to study the early matter-dominated epoch.

\subsection{Model II: p-wave scattered fermion dark matter}
\label{fermionDM}

Now we consider a second model featuring a fermionic DM $\psi_\chi$ which interacts with fermionic bath particles $\psi_\gamma$ through a scalar mediator $\varphi_M$ by Yukawa interaction $y \, \bar\psi_\chi\psi_\gamma\varphi_M$. We again emphasize that $\psi_\gamma$ can be either a Standard Model fermion or any other BSM fermion in the thermal bath. In this case, the elastic scattering processes $\psi_\chi\psi_\gamma\to\psi_\chi\psi_\gamma$, $\psi_\chi\bar{\psi}_\gamma\to\psi_\chi\bar{\psi}_\gamma$ and their conjugates undergo via $p$-wave amplitudes. The rate of momentum exchange is given by
\begin{eqnarray}
	\gamma_{\rm el}(T) = \frac{341}{756}\pi^{3}\,y^{4}\frac{m_{\chi}^{3}}{(M - m_{\chi})^{2}}\left(\frac{T}{m_{\chi}}\right)^{6}, \quad  {\rm for} \quad m_\chi < M \leq 2m_\chi\,,
\end{eqnarray}
where $M$ denotes the mass of the scalar mediator $\varphi_M$, and we neglect the mass of $\psi_\gamma$.
The total annihilation cross-section for the processes $\psi_\chi\bar{\psi}_\chi\to \psi_\gamma\bar{\psi}_\gamma$, and $\psi_\chi\psi_\chi\to \psi_\gamma\psi_\gamma$, responsible for freeze-out mechanism, are given by
\begin{align}
	\sigma_{\rm ann} = \frac{y^{4}}{16\pi s} \sqrt{\frac{s}{s-4m_{\chi}^{2}}}\left[\frac{7\Delta m^{4} + 4M^{2}s}{\Delta m^{4} + M^{2}s} - 4\,\frac{7\Delta m^{4} + 4M^{2}s - 2m_{\chi}^{2}s}{s(s-4m_{\chi}^{2})\tau(s)} \coth^{-1}\tau(s)\right]\,, 
\end{align}
where we define
\begin{align}
    \Delta m^{2} \equiv M^2-m_\chi^2\,, \quad {\rm and} \quad \tau(s) \equiv \frac{s-2m_{\chi}^{2} + 2M^{2}}{\sqrt{s(s-4m_{\chi}^{2})}}\,.
\end{align}
\begin{figure}[t!]
     \centering
      \includegraphics[width=0.8\textwidth]{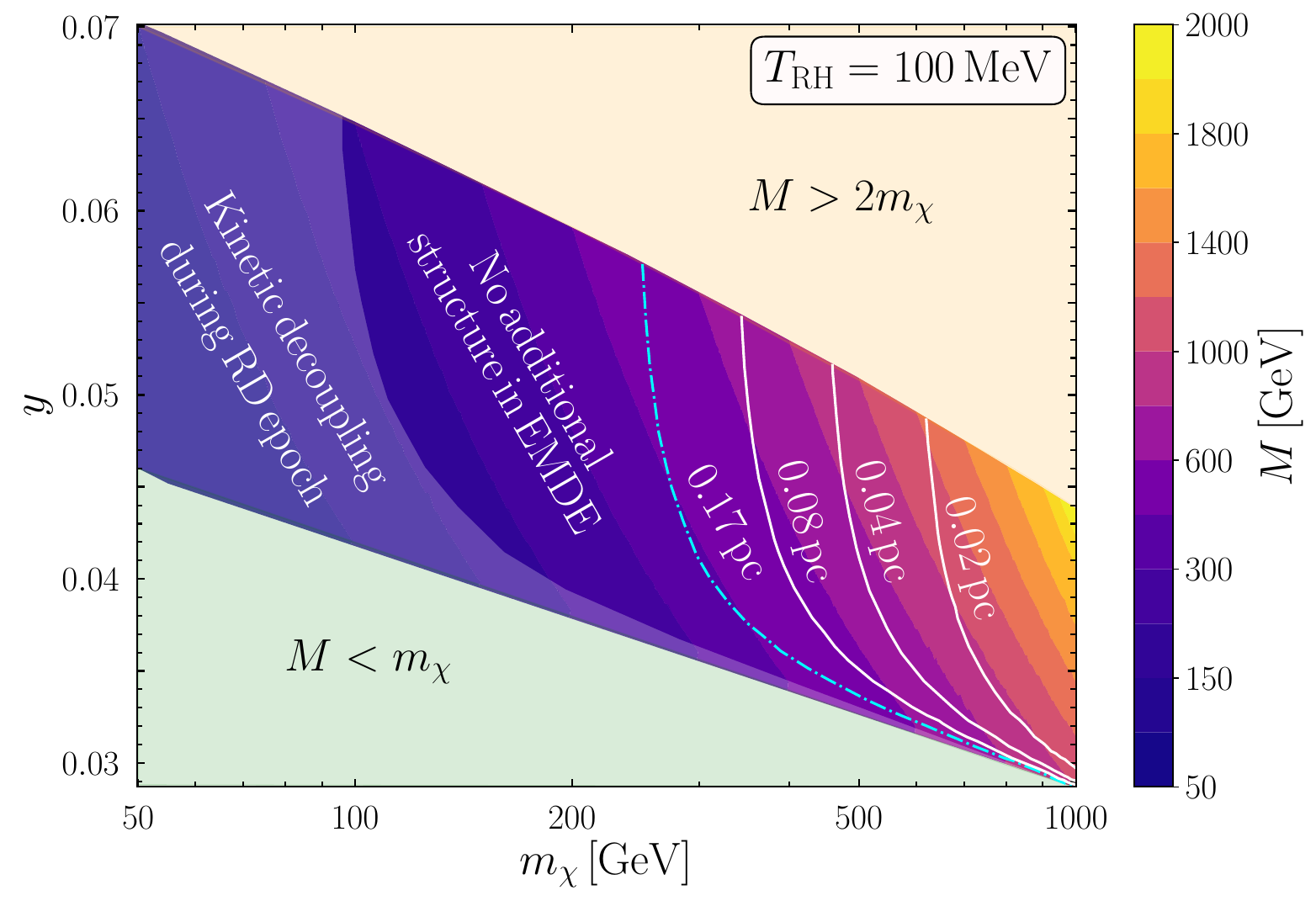}
     \caption{\sf\it The allowed parameter region satisfying  $m_\chi<M\leq 2m_\chi$ and reproducing the present day relic density of dark matter for a fixed value of $T_{\rm RH} = 100$ MeV. In the dark-shaded region, dark matter kinetically decouples during the EMDE, while it decouples during radiation-dominated epoch in the light-shaded region. Enhancement in the sub-earth halo population can not be observed in the region on the left side of the cyan dashed curve, even though DM remains fully kinetically decoupled during EMDE. The isocontours of the free-streaming horizon are shown by white lines.}
        \label{fig:point_yukawa}
\end{figure}

In Fig.~\ref{fig:point_yukawa}, we present the allowed parameter space within $m_\chi < M\leq 2m_\chi$,  satisfying the present day relic density of dark matter, for a fixed value of reheating temperature $T_{\rm RH}=100$ MeV. Salient features of this model illustrated in Fig.~\ref{fig:point_yukawa} are described below: 
\begin{itemize}  

\item The DM shows full kinetic decoupling during the EMDE in the dark-coloured area, while the relic is satisfied in the entire parameter space within the specified boundaries. 

\item The isocontours of the free-streaming horizon are projected as white curves on the allowed parameter space. Due to the full decoupling of the DM, the value of $\lambda_{\rm fsh}$ is approximately two orders of magnitude smaller compared to a similarly situated benchmark point in Fig.~\ref{figure_point}, where the DM decouples only partially.

\item The region left to the cyan dot-dashed line does not exhibit a boost in the population of sub-earth halos even if DM becomes fully decoupled during the EMDE. This is because the free-streaming horizon of the DM in that region exceeds the cut-off for the corresponding $T_{\rm RH}$. 
\end{itemize}

Analogous to the previous model in this case as well, the annihilation cross-section required to produce the same amount of DM relic density in an EMDE with $\Gamma_\phi\sim T$ lies in between that required in $\Gamma_\phi\sim {\rm const.}$ scenario and purely radiation dominated universe. Moreover, for $\Gamma_\phi\sim {\rm const.}$, $p$-wave elastic scattering is able to partially decouple the DM during EMDE, whereas the DM gets fully kinetically decoupled in EMDE with $\Gamma_\phi\sim T$, resulting in a comparatively reduced free-streaming horizon of the DM, provided the decoupling temperature is same. Therefore, the boost in the sub-earth halo population is expected to be more in the $\Gamma_\phi\sim T$, putting it in an advantageous situation in terms of observational probes.  

\begin{figure}[t!]
     \centering
      \includegraphics[width=0.65\textwidth]{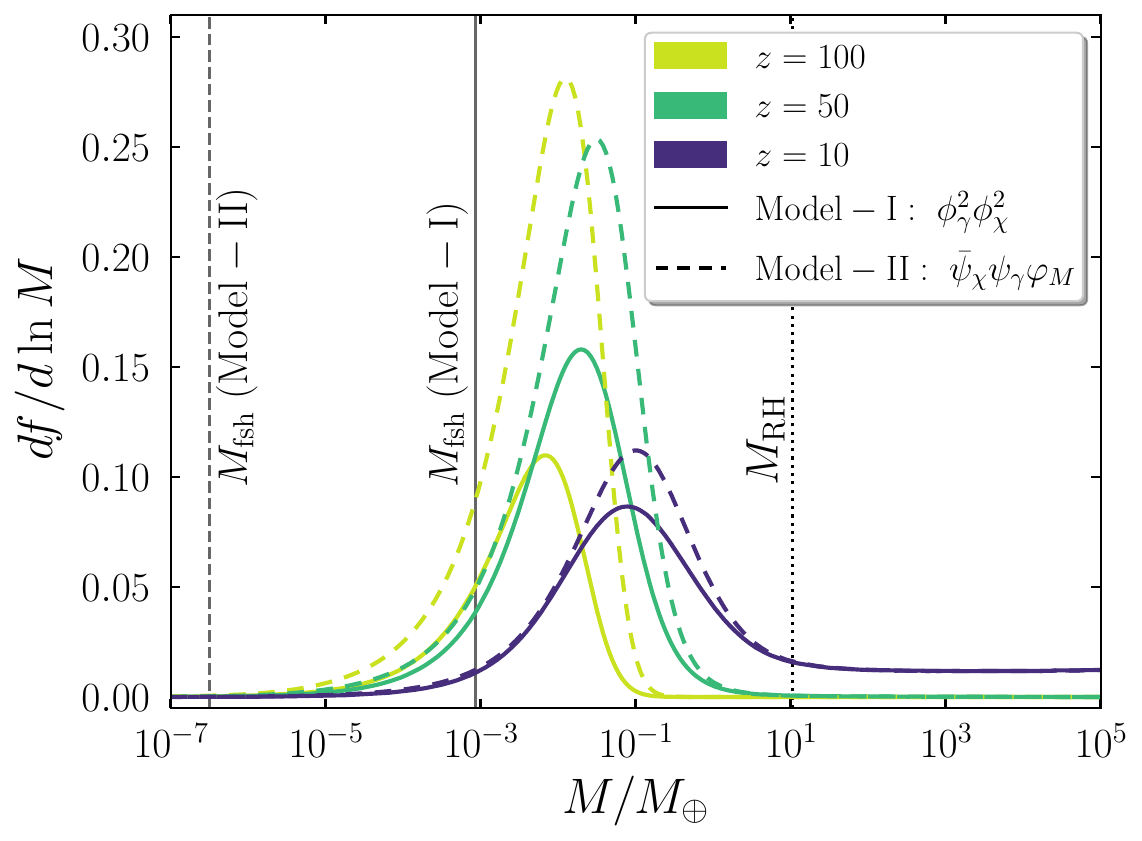}
     \caption{\sf\it Differential halo mass fraction is demonstrated for both the models at different redshift values. For the benchmark points as mentioned in the caption of Fig.~\ref{fig:perturb}, the free-streaming horizon of DM turns out to be $0.264$ $pc$ and $0.019$ $pc$ for Model I and Model II, respectively. The differential halo mass function peaks around $M = 0.1\,M_\oplus$.}
        \label{fig:structure_yukawa}
\end{figure}

We illustrate the differential halo mass function at several redshift values for both Model I and II in Fig.~\ref{fig:structure_yukawa}, keeping $T_{\rm RH}=10$ MeV and $m_\chi=95$ GeV for both cases. The relevant couplings are chosen, as in Fig.~\ref{fig:perturb}, to satisfy the present-day relic density. For this benchmark point, the DM decouples around $T=2.2$ GeV for both Models. However, the free-streaming horizon of the DM is $\lambda_{\rm fsh}=0.264$ pc for Model I and $0.019$ pc for Model II, respectively. The smaller $\lambda_{\rm fsh}$ in Model II is attributed to the extra cooling due to full kinetic decoupling before the end of reheating. The additional cooling in Model II is further confirmed by the DM temperature at $a_{\rm RH}$, with $T_{\chi}(a_{\rm RH})$ being $55.5$ KeV for Model I and $12$ eV for Model II. As shown in Fig.~\ref{fig:structure_yukawa}, the differential halo mass function peaks around $M = 0.1\,M_\oplus$ for both models, while the absolute value is higher in Model II due to the reduced $\lambda_{\rm fsh}$.

\section{Conclusions}
\label{conc}

We have comprehensively studied the chemical and kinetic decoupling of thermal dark matter in the presence of a pre-BBN early matter-dominated era. Specifically, we explore a scenario where a long-lived scalar field dominated the energy density of the universe, eventually transferring energy to the radiation through a temperature-dependent dissipation rate ($\Gamma_\phi \sim T$). We demonstrate that the kinetic decoupling of the DM during the EMDE is partial if the elastic scattering of the DM with the bath particles is dominated by $s$-wave partial wave amplitudes, while the DM fully decouples during the EMDE if it undergoes $p$-wave scattering. This is in contrast to an EMDE with a constant rate of energy dissipation ($\Gamma_\phi\sim {\rm const.}$), where $s$-wave scatterings are inefficient to kinetically decouple the DM, while $p$-wave scatterings can partially decouple it from the thermal bath. 

An early kinetic decoupling reduces the free-streaming horizon of the DM ($\lambda_{\rm fsh}$) due to additional cooling of the DM during the EMDE. In addition, the presence of an EMDE also enhances the matter perturbations for scales entering the horizon during this epoch. The population of the DM halos with masses smaller than $M_{\rm RH}$ receive a boost due to the combined effect of reduced free-streaming horizon and enhanced matter perturbations. The boost in the number of small-scale DM halos may in turn lead to an increase in the gamma-ray annihilation signals from the DM halos, offering an interesting probe for the early matter-dominated cosmology. Gravitational methods, such as pulsar timing arrays, microlensing, and detection of primordial gravitational waves provide alternative prospects to probe the physics of the early matter-dominated universe.

However, kinetic decoupling of the DM during the EMDE does not guarantee an enhanced structure growth. If the annihilation of the DM which determines its relic and the elastic scattering with bath particles are governed by the same interaction strength, the free-streaming horizon reduces in the EMDE. On the other hand, in the presence of multiple interactions which decorrelate the mechanisms for chemical and kinetic decoupling, the free-streaming horizon may not be reduced in the presence of the EMDE unless the kinetic decoupling temperature is greater than a threshold value compared to the reheating temperature, see Fig.~\ref{fig:free_stream}.  

Moreover, we have shown that the formation of DM halos receives a boost iff the free-streaming horizon is smaller than a certain cut-off ($\lambda^c_{\rm fsh}$). We have quantified $\lambda^c_{\rm fsh}$, which depends on the reheating temperature, by comparing $\sigma(M)$ in the presence of EMDE and in the purely radiation-dominated universe with $\lambda_{\rm fsh}=0$. 

As mentioned above, if the chemical and kinetic decoupling are correlated by the same interaction strengths, the requirement of kinetic decoupling and the cut-off $\lambda^c_{\rm fsh}$, ensuring enhancement in small-scale structure, provide a new handle to distinguish the parameter space of a DM model in presence of a non-standard cosmological epoch. We utilize this idea to identify novel areas in the parameter space of two simplified models: one with a scalar DM involving $s$-wave scattering by scalar quartic interactions (Model I), and the other with fermionic DM showing $p$-wave scattering by Yukawa interactions (Model II). We obtain the viable region of parameter space in the WIMP DM mass range for both models, where the DM decouples partially (Model I) or fully (Model II) resulting in $\lambda_{\rm fsh}<\lambda^c_{\rm fsh}$, during an EMDE with temperature-dependent entropy injection which yields a sub-GeV reheating temperature. This results in an amplification of the sub-earth halo population with a peak around $M=0.1M_\oplus$ for $T_{\rm RH} = 10$ MeV. In comparison, an EMDE with $\Gamma_\phi\sim {\rm const.}$ requires larger values of $T_{\rm RH}\sim {\rm GeV}$ to reproduce the correct relic density of the DM, which would result in the formation of structures at much smaller scales($M < 10^{-5}M_\oplus$) compared to $M=0.1M_\oplus$, thus providing a direct observational access to the future experiments in probing different non-standard cosmological epochs. 

\acknowledgments
 A.B. acknowledges support from the Department of Atomic Energy, Govt.\,of India. A.B. would like to express special thanks to the Knut and Alice Wallenberg foundation (grant KAW 2017.0100, SHIFT project) and the Chalmers University of Technology, G\"oteborg, Sweden for support during the initial stages of this work. This research of D.C. is supported by an initiation grant IITK/PHY/2019413 at IIT Kanpur and by DST-SERB grants SERB/CRG/2021/007579 and SERB/CRG/2022/001922. A.H. and M.S.I would like to thank the MHRD, Govt. of India for the research fellowship.

\appendix

\section{Perturbation equations}
\label{sec:pert_eq}

We provide the detailed equations governing the density and velocity perturbations of the scalar $\phi$, radiation, and the DM fluids below.
\begin{align}
    \label{perturbeq}
        & \delta_{\phi}^\prime +\frac{\theta_{\phi}}{{a^2H}}-3\Phi^\prime = -\frac{\Gamma_{\phi}}{{aH}}\left[\Phi+\frac{\delta\Gamma_\phi}{\Gamma_\phi}\right] \\
        \label{deltadm}
        & \delta_{\chi}^\prime +\frac{\theta_{\chi}}{{a^2H}}-3\Phi^\prime =  \Sigma^{\delta_\chi}_{\rm ann}\\
         \label{deltar}
        & \delta_{{\gamma}}^\prime +\frac{4}{3}\frac{\theta_{{\gamma}}}{{a^2H}}-4\Phi^\prime = \frac{\Gamma_{\phi}\rho_{\phi}}{{aH}\rho_{{\gamma}}}\left[\delta_{\phi}-\delta_{\gamma}+\Phi+\frac{\delta\Gamma_\phi}{\Gamma_\phi}\right] + \Sigma^{\delta_\gamma}_{\rm ann} \\
        \label{thetascalar}
        & \theta_{\phi}^\prime+ \frac{\theta_{\phi}}{a}-\frac{k^2}{{a^2H}}\Phi = 0 \\
        \label{thetadm}
        & \theta_{\chi}^\prime + \frac{\theta_{\chi}}{a}-\frac{k^2}{{a^2H}}\Phi = 0 \\
        \label{thetar}
        & \theta_{\gamma}^\prime - \frac{k^2}{{a^2H}}\left(\frac{\delta_{\gamma}}{4}+\Phi\right) = \frac{\Gamma_{\phi}}{{aH}}\frac{\rho_{\phi}}{\rho_{{\gamma}}}\left[ \frac{3}{4}\theta_{\phi}-\theta_{{\gamma}}\right] + \Sigma^{\theta_\gamma}_{\rm ann}\,,
\end{align}
where the annihilation terms are given by
\begin{align}
    \Sigma^{\delta_\chi}_{\rm ann} & = -\frac{\langle \sigma v\rangle_{\rm ann}}{{m_{\chi}aH}}\left[\rho_{\chi}(\delta_{\chi}+\Phi)-\frac{\rho_{\chi,{\rm eq}}^2}{\rho_{\chi}}(2\delta_{\chi,{\rm eq}}-\delta_{\chi}+\Phi)\right]-\frac{d\langle\sigma v\rangle_{\rm ann}}{dT}\frac{T \delta_\gamma}{4m_{\chi}aH}\frac{\rho_{\chi}^2-\rho_{\chi,{\rm eq}}^2}{\rho_{\chi}}\,, \\
    \Sigma^{\delta_\gamma}_{\rm ann} & = \frac{\langle \sigma v\rangle_{\rm ann}}{{m_{\chi}aH}}\left[\frac{\rho_{\chi}^2}{\rho_{{\gamma}}}(2\delta_{\chi}-\delta_{{\gamma}}+\Phi) -\frac{\rho_{\chi,{\rm eq}}^2}{\rho_{{\gamma}}}(2\delta_{\chi,{\rm eq}}-\delta_{{\gamma}}+\Phi)\right] +\frac{d\langle\sigma v\rangle_{\rm ann}}{dT}\frac{T \delta_\gamma}{4m_{\chi}aH}\frac{\rho_{\chi}^2-\rho_{\chi,{\rm eq}}^2}{\rho_{\gamma}}\,, \\
    \Sigma^{\theta_\gamma}_{\rm ann} & = \frac{\langle\sigma v\rangle_{\rm ann}}{{m_{\chi}aH}}\frac{\rho^2_{\chi}-\rho_{\chi,{\rm eq}}^2}{\rho_{{\gamma}}}\left[\frac{3}{4}\theta_{\chi}-\theta_{{\gamma}}\right]\,.
\end{align}

The Eqs.(\ref{perturbeq})-(\ref{thetar}) are closed by the time-time
component of the perturbed Einstein equation as \cite{Fan:2014zua}
\begin{eqnarray}
        \label{phi}
         a\Phi^\prime = -\left(\frac{k^2}{3{a^2H^2}}+1\right)\Phi-\frac{1}{6{H^2M_{\rm pl}^2}}\left(\rho_{\phi}\delta_{\phi}+\rho_{{\gamma}}\delta_{{\gamma}}+\rho_{\chi}\delta_{\chi}\right).
    \end{eqnarray}
Here, the prime denotes differentiation with respect to the scale factor ($a$), and $\rho_i$'s denotes the background densities. The above equations are written in the Fourier space where the $\nabla^2$ is replaced by $-k^2$.

For EMDE with temperature-dependent entropy injection, i.e., $\Gamma_\phi\sim T$, we get $\delta\Gamma_\phi/\Gamma_\phi \sim \delta_\gamma/4$. The $\langle\sigma v\rangle_{\rm ann}$ and its derivative with respect to the temperature has to be calculated for a given DM model. For example, in the Model I of Sec.~\ref{scalarDM}, we find
\begin{align}
    \frac{d\langle\sigma v\rangle_{\rm ann}}{dT} = - \frac{3\lambda^2}{32\pi m_\chi^3}\,.
\end{align}

\subsection*{Fitting $\delta_\chi$ with empirical function}

\begin{figure}[t!]
     \centering
      \includegraphics[width=0.65\textwidth]
      {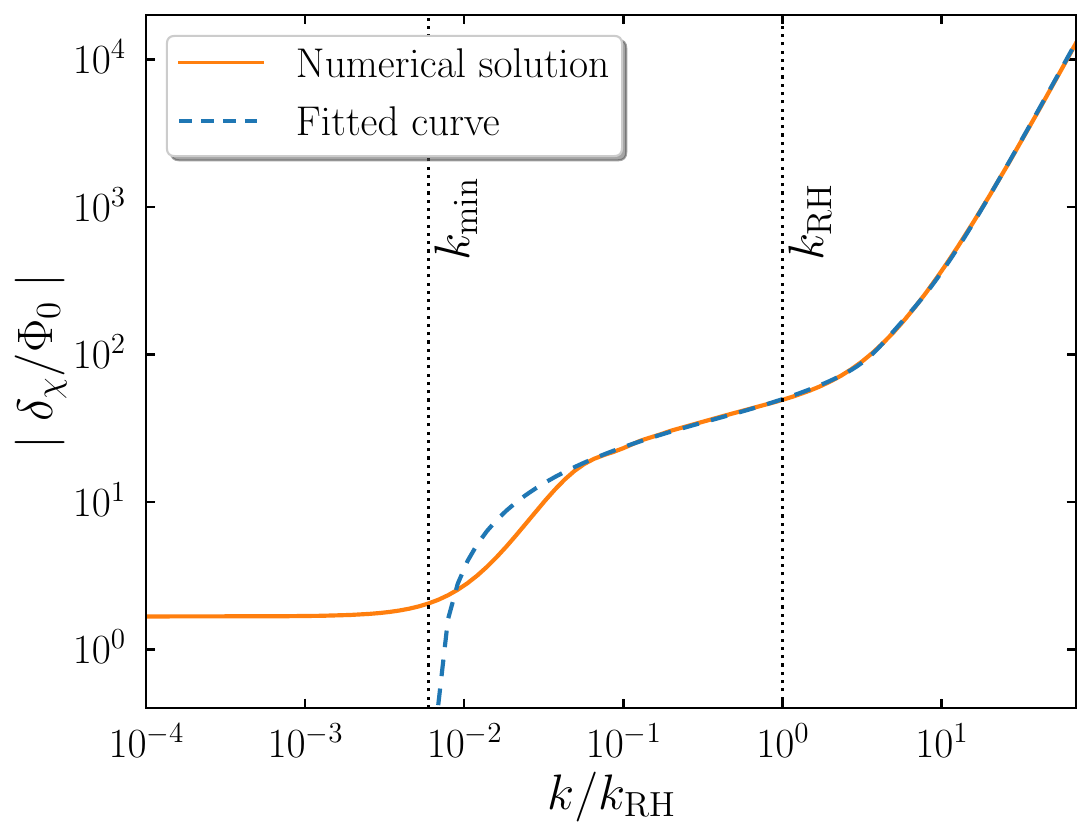}
     \caption{\sf\it Full numerical solution (orange) and best-fit curve (blue dashed) for $\delta_{\chi}$ as a function of $ k/k_{\rm RH}$ are shown at $a/a_{\rm RH} = 192.5$ keeping $T_{\rm RH} = 10$ MeV. For $k < k_{\rm min}$, $\delta_\chi$ is constant as these modes are still super-Hubble, for $k_{\rm min} < k < 0.1\,k_{\rm RH}$ $\delta_\chi$ grows since the metric perturbation $\Phi$ has not decayed, while for $0.1\,k_{\rm RH} < k < k_{\rm RH}$, $\delta_\chi$ settles into logarithmic growth. Modes with $k > k_{\rm RH}$ entered the horizon during the EMDE, resulting in $\delta_\chi$ displaying power-law momentum dependence. The best-fit curve closely follows the full numerical solution for $k> 0.1\,k_{\rm RH}$.}
        \label{fig:fit}
\end{figure}
We use the following empirical function to fit the full numerical solution of $\delta_\chi(a)$:
\begin{align}  
    \label{fit function_1}
    \delta_{\chi} = \frac{10}{9}C_1(k/k_{\rm RH})\Phi_0\ln\Bigl(C_2(k/k_{\rm RH})\frac{a}{a_{\rm hor}}\Bigr)\,,
\end{align}
where the functions $C_1(x)$ and $C_2(x)$, with $x\equiv k/k_{\rm RH}$, are defined in terms of unknown coefficients $\alpha_i$ as
\begin{align}
    \label{fit function_2}
    \nonumber
   & C_1(x) = \exp\left[\frac{\alpha_1}{\left(1+\alpha_2\left(\ln(x)-\alpha_3\right)^2\right)^{\alpha_4}}\right]
    \left[9.11 e^{\alpha_5} S\left(\alpha_6-x\right) + \frac{3}{5} \alpha_7 x^2 S\left(x-\alpha_6\right)\right], \\   
    & \ln C_2(x) = \ln(0.594) S\left(\alpha_6-x\right) + \ln\left(\frac{e}{x^2}\right) S\left(x-\alpha_6\right)\,,~ \, S(y) = \frac{1}{2}\left[\tanh\left(\frac{y}{2}\right)+1\right].
\end{align}
The coefficients $\alpha_i$ are obtained by fitting the above empirical expression with the numerical solution of $\delta_\chi$ for $T_{\rm RH} = 10$ MeV, and for the modes $k> 0.05\, k_{\rm RH}$, as given below
\begin{align*}\begin{array}{llllllll}
     \alpha_1 = 0.858\pm0.039, && \alpha_2 = 2.533\pm0.287, && \alpha_3 = 1.455\pm0.025, && \alpha_4 = 0.791\pm0.095,\\
    \alpha_5 = -0.0974\pm0.023, && \alpha_6 = 3.225\pm0.153, && \alpha_7 = 0.647\pm0.015\,. &&
\end{array}
\end{align*}
In Fig.~\ref{fig:fit} we show that the best-fit curve (in orange) matches well with the numerical solution (blue dashed curve) for $k> 0.05 \,k_{\rm RH}$.

\bibliographystyle{JHEP}
\bibliography{main}

\end{document}